\documentclass[a4paper,11pt]{article}

\usepackage{jcappub}

\usepackage[T1]{fontenc} 

\usepackage{epsfig}
\usepackage{latexsym}
\usepackage{graphicx}
\usepackage{epstopdf} 
\usepackage[justification=RaggedRight]{caption}
\usepackage{amssymb}
\usepackage{subfig}
\usepackage{color}
\usepackage{multirow}
\usepackage{color}
\usepackage{rotating}
\usepackage{ifthen}
\usepackage{epsfig}
\usepackage{booktabs} 
\usepackage{hyperref} 
\usepackage{hyperref}
\hypersetup{
    colorlinks,%
    citecolor=blue,%
    filecolor=blue,%
    linkcolor=blue,%
    urlcolor=blue
}
\def\be{\begin{equation}}
\def\ee{\end{equation}}

\newcommand{\bwt}{\begin{widetext}}
\newcommand{\ewt}{\end{widetext}}
\newcommand{\bdm}{\begin{displaymath}}
\newcommand{\edm}{\end{displaymath}}
\newcommand{\bea}{\begin{eqnarray}}
\newcommand{\eea}{\end{eqnarray}}

\newcommand{\beqra}{\begin{eqnarray}}
\newcommand{\eeqra}{\end{eqnarray}}
\newcommand{\beq}{\begin{equation}}
\newcommand{\eeq}{\end{equation}}

\title{LHC prospects for minimal decaying Dark Matter}

\author{Giorgio Arcadi,}

\author{Laura Covi}

\author{and Federico Dradi}

\affiliation{Institute for Theoretical Physics, 
Georg-August University G\"ottingen, 
Friedrich-Hund-Platz~1, G\"ottingen, D-37077 Germany}

\abstract{
We study the possible signals at LHC of the minimal models of
decaying dark matter. Those models are characterized by the fact
that DM interacts with SM particles through renormalizable coupling 
with an additional heavier charged state. Such interaction 
allows to produce a substantial abundance of DM in the early
Universe via the decay of the charged heavy state, either in- 
or out-of-equilibrium. Moreover additional couplings of the
charged particle open up decay channels for the DM, which can
nevertheless be sufficiently long-lived to be a good DM 
candidate and within reach of future Indirect Detection observations.
We compare the cosmologically favored parameter regions to the
LHC discovery reach and discuss the possibility of simultaneous 
detection of DM decay in Indirect Detection.
}

\begin{document} 
\maketitle
\flushbottom

\section{Introduction}

After the first run of the LHC, no signal of New Physics has 
been found yet in the traditionally DM-motivated channels containing
missing energy \cite{ATLAS-CONF-2012-147,CMS-PAS-EXO-12-048}. 
While the next LHC run could still bring
a WIMP-like signal, especially from the electroweakly charged 
sector, we would like here to consider a different type of scenario, 
connected to the cosmologically well-motivated case of decaying 
Dark Matter.

In fact, it is perfectly possible that the DM particle is not
stable, but just very long-lived and in such case immediately
two usual DM assumptions are lifted:
a) no symmetry is introduced to justify the stability
of Dark Matter and therefore the particle can interact also
as single state with charged SM and non-SM states;
b) to ensure survival until the present age, the DM particle
has to interact with suppressed couplings to any sector and
therefore can be naturally produced not through the WIMP
mechanism, which relies on electroweak-size cross-section,
but via the FIMP or SuperWIMP mechanisms, based on much
smaller couplings.

A very minimal setting of this type has been recently 
proposed in~\cite{Arcadi:2013aba}, featuring a DM Maiorana fermion 
and a scalar charged under the SM gauge group.
It is minimal in the sense that just a minimal particle content has 
been added to the SM and that the additional couplings are all 
renormalizable. Such a scheme can be embedded in more complex
models, like supersymmetry with R-parity violation, but the main 
phenomenological characteristics are independent 
from the particle physics framework, at least for what concerns 
DM phenomenology. 
We will thus here concentrate on such minimal constructions as the 
more conservative case, since in the presence of a larger field 
content or more couplings, additional signals could be
accessible at LHC or in DM indirect detection.

Due to the long lifetime of the DM, requiring at least one of 
the additional charged particles to be within the LHC reach 
constrains all couplings to be suppressed and therefore points
also to regions of (near)-collider-metastability for these
exotic states. It is therefore natural to look for such particles 
at the LHC in the presence of displaced vertices or metastable tracks (MP).
These channels have recently attracted more 
attention~\cite{Han:2007ae,Ishiwata:2008tp,Chang:2009sv,Meade:2009mu,Meade:2010ji,
Bobrovskyi:2011vx,Heisig:2011dr,Ghosh:2012pq,Heisig:2012zq,Graham:2012th,
Heisig:2013rya,Cirelli:2014dsa}
and are indispensable in order to cover a wide range of
cosmologically interesting scenarios~\cite{Feng:2004yi,Hirsch:2005ag,ArkaniHamed:2008qp,
Hall:2009bx,Cheung:2010gk,Covi:2010au,Endo:2010ya,Lindert:2011td} like the one presented 
here.

This paper is organized as follows: we will discuss the type
of minimal models we study in section 2 together with the
connection to DM indirect detection and the DM production
mechanisms. We will then study in section 3 the LHC phenomenology in 
the two scenarios of scalar field carrying color or only electroweak
charge. Section 4 will be then devoted to the discussion of the results 
while our conclusions will be given in Section 5.

\section{The model}

We consider here the minimal model introduced in~\cite{Arcadi:2013aba} 
(see also~\cite{Garny:2010eg,Garny:2011ii,Garny:2012vt} for simular setups) 
featuring a Maiorana fermion, singlet with respect to the SM gauge group and
dark matter candidate, and a single scalar field multiplet $\Sigma_f$, 
non-trivially charged under at least one of the SM gauge groups.
We assume that these fields interact among themselves and with the
SM only via renormalizable Yukawa-type couplings according to the quantum 
number chosen for $\Sigma_f$. In particular the interaction between
the DM and $\Sigma_f$ will be of the form:
\begin{equation}
L_{\rm eff} =\lambda_{\psi f}\;\bar \psi f \;\Sigma_{f}^{\dagger} +h.c.
\label{eq:DMint}
\end{equation}  
where $\psi$ is the DM Majorana fermion, $f$ any chiral SM fermion and
$\Sigma_{f}$ denotes a scalar with quantum numbers equal to $f$.

No symmetry is imposed to guarantee the stability of the DM, allowing 
couplings of $\Sigma_{f}$ with only SM fermions. 
Depending on the quantum numbers a rather broad variety of operators may arise: 
\begin{align}
\label{eq:lagrangian}
L_{\rm eff}&= \lambda_{1q}  \bar d \ell \Sigma_q+h.c. 
   &\Sigma_q=(3,2,1/3) \nonumber\\
L_{\rm eff} &= \lambda_{1u} \bar{d} d^c \Sigma_u^\dagger+h.c.
 & \Sigma_u=(3,1,4/3) \nonumber\\ 
L_{\rm eff} &= \lambda_{1d}  \bar q \ell^c \Sigma_d +
\lambda_{2d} \bar u d^c \Sigma_d^{\dagger} +\lambda_{3d} \bar u e^c \Sigma_d + 
\lambda_{4d} \bar q q^c \Sigma_d^{\dagger}+h.c. 
 &  \Sigma_d=(3,1,-2/3) \nonumber\\
L_{\rm eff} &= {\lambda}_{1\ell} \bar e \ell \Sigma_\ell + \lambda_{2\ell} 
\bar d q \Sigma_\ell+\lambda_{3\ell} \bar u q \tilde{\Sigma_\ell}+h.c.
 & \Sigma_\ell=(1,2,-1) \nonumber\\
L_{\rm eff} &=  \lambda_{1e} \bar \ell \ell^c \Sigma_e+h.c.
& \Sigma_e=(1,1,-2)
\end{align}
where $q,\ell $ denote the SM $ SU(2)_L$ LH doublets, while $u, d, e $ are RH $SU(2)_L$ singlets, 
and the superscript $^c$ indicates the charge-conjugated field, $f^c = C \bar{f}^t $ while 
$\tilde{\Sigma}_f \equiv i \sigma_2 \Sigma_f^{*}$. On the right the quantum numbers of the 
$\Sigma_f $ fields are specified according to the SM gauge groups $ SU(3)_c \times SU(2)_L 
\times U(1)_Y $. 
We are here suppressing flavour indices, even if some couplings like $ \bar q q^c, \bar\ell \ell^c$ 
must be antisymmetric in flavour and vanish for a single generation, and considering in each 
line the presence of a single scalar field $\Sigma_f$. 
We have then that the new particle sector can just be described by two mass scales $m_\psi$ 
and $ m_\Sigma$ and a few Yukawa couplings.  
The scalar field $\Sigma_f$ is also coupled, according to its assignment of quantum numbers, 
given above, to the SM group gauge bosons.

The interactions~(\ref{eq:DMint}) and (\ref{eq:lagrangian}) induce three-body decays for the Dark Matter
with a rate given by, up to kinematical and multiplicity factors:
\begin{equation}
\Gamma_{\rm DM}=\frac{c_f |\lambda_{\psi f}|^2 |\overline{\lambda|^2}}{384 {\left(2 
\pi\right)}^3}\frac{m_\psi^5}{m_{\Sigma_f}^4}, \,\,\,\overline{\lambda}=\lambda_{if}
,\,\,\,\,\,i=1,\cdots,4,\,\,\,f=q,u,d,\ell,e
\label{eq:Gamma_DM}
\end{equation}
with $c_f = $ counting the number of degrees of freedom of the intermediate $\Sigma_f$ 
($c_f = 6, 3, 3, 2, 1 $ for $\Sigma_q,  \Sigma_u, \Sigma_d, \Sigma_\ell, \Sigma_e$ respectively).
The decay channels can be different depending on the quantum numbers of the intermediate
particle $\Sigma_f$.
We distinguish substantially four types of decay channels:
\begin{eqnarray}
\psi \rightarrow & & \overline{u}u\nu, \overline{d}d\nu \quad \mbox{for} \quad \Sigma_d, 
\Sigma_\ell \nonumber\\
\psi \rightarrow & & \overline{u}d l \quad \mbox{for} \quad \Sigma_q, \Sigma_d \nonumber\\
\psi \rightarrow & & udd \quad \mbox{for} \quad \Sigma_u, \Sigma_d \nonumber\\
\psi \rightarrow & & l \overline{l} \nu \quad \mbox{for} \quad \Sigma_\ell, \Sigma_e
\end{eqnarray} 
For simplicity we focus on signals of the type $f\overline{f}\nu$ with $f$ being either a 
quark or a charged lepton of any generation, denoted by $l$.  
We will also assume that the decays of DM are flavour conserving. If this is not the case 
bounds from flavour violation decays of mesons and leptons arise (see e.g.~\cite{Barbier:2004ez} 
for a complete list). However they are sensitively weaker than the one imposed by cosmology and 
DM ID. We will also assume that, for a given assignment of the quantum numbers of $\Sigma_f$, 
only one of the allowed operators, as reported in~(\ref{eq:lagrangian}),
dominates. We just comment that some of the effective lagrangians ~(\ref{eq:lagrangian}) violate 
both lepton and baryon number and in case of contemporary presence of lepton and baryon number 
violating operators, very strong constraints from the stability of the proton 
arise~\cite{Smirnov:1996bg}. We will neglect here this possibility.  

All the considered DM decay channels are already severely constrained from Indirect Detection. 
The hadronic decay channels, namely $d\overline{d}\nu$ and $u\overline{u}\nu$ are mostly 
constrained by antiproton searches, which give bounds on the DM lifetime varying between 
$ 10^{26}-10^{28}$ s for $m_\psi > 100 $ GeV~\cite{Garny:2012vt} and can become as stringent 
as $10^{29}\mbox{s}$ for values of the DM mass down to 1 GeV~\cite{Fornengo:2013xda}.
In the case of the leptonic decays, comparably severe constraints, ranging from approximately 
$10^{27}$s to $10^{29}$s, according to the decay channel, for DM masses between 10 GeV and 2 TeV,
are obtained by the recent measurements by AMS of the positron flux and positron 
fraction~\cite{Ibarra:2013zia}. 
In the case of the $\Sigma_\ell $ mediator, one-loop induced decay processes in $Z\gamma$ 
and $\nu\gamma$ may also be important. In particular the latter can originate monochromatic 
$\gamma$-ray lines. Current searches give bounds which can be as strong as 
$10^{29 \div 30}$ seconds for 
$m_\psi=1-10\,\mbox{GeV}$~\cite{Garny:2010eg,Ackermann:2012qk,Ackermann:2013uma}. 
All these bounds can be satisfied only for a very small value of the product of the couplings,
namely $\lambda_{\psi\,f}\,\bar{\lambda}\lesssim 10^{-(16 \div 22)}$, for masses of the scalar 
field $\Sigma_f $ within the kinematical reach of the LHC. 

While the present lower limits on the DM lifetime only constrain the product of the
couplings, the DM production mechanisms in the early universe also provide bounds
on the ratio of the couplings. In fact, in this kind of models, the correct amount 
of DM relic density can be generated from the decays of the field $\Sigma_f$ either 
in thermal equilibrium (freeze-in 
production~\cite{Hall:2009bx,Chu:2011be,Chu:2013jja,Klasen:2013ypa,Blennow:2013jba}) or 
out-of-equilibrium (SuperWIMP 
production~\cite{Covi:1999ty,Feng:2003xh,Feng:2003uy})~\footnote{It is as well possible, 
alternatively, that the DM acquires its relic density through the WIMP paradigm. However 
this setup corresponds to the simple scenario with $\lambda_{\psi f} \sim O(1)$ and an 
irrelevant coupling of $\Sigma_f$ to SM fields. We will not consider this possibility in 
this work (see instead e.g.~\cite{Gomez:2014lva} for a scenario of this kind with DM coupling 
to the top quark).}. 

Since the two processes take place at quite different cosmological epochs,
the DM relic density can expressed, in very good approximation, as the sum of
the two contributions \cite{Arcadi:2013aba}:
\begin{align}
\label{Omegah2-DM2}
& \Omega_{\rm DM}h^2=\Omega^{\rm FI}_{\rm DM}h^2+\Omega_{\rm DM}^{\rm SW}h^2\nonumber\\
& \approx 1.09\times 10^{27}  \frac{g_\Sigma}{g_{*}^{3/2}} \frac{x \Gamma\left(\Sigma_f 
\rightarrow f DM\right)}{m_{\Sigma}}+x Br\left(\Sigma_f \rightarrow f DM\right) 
\Omega_\Sigma h^2 \nonumber\\
& \approx x Br\left(\Sigma_f \rightarrow f DM\right) 
 \left[ 0.717 \frac{ g_\Sigma}{g_{*}^{3/2}} \left(\frac{1 \mbox{s}}{\tau_\Sigma}\right) 
\left(\frac{1 \mbox{TeV}}{ m_{\Sigma}}\right) +  \Omega_\Sigma h^2 \right]
\end{align}
where $x=\frac{m_{\psi}}{m_{\Sigma}}$, $g_\Sigma$ and $g_{*}$ are, respectively, the internal 
degrees of freedom of the field $\Sigma_f$ and the relativistic degrees of freedom of the 
primordial plasma  at the time of DM production, $\tau_\Sigma$ is the lifetime of the $\Sigma_f$ 
field defined, up to kinematical factors, as:
\begin{equation}
\label{eq:rate1}
\tau_\Sigma^{-1}=\Gamma_\Sigma=\frac{|\lambda_{\psi\,f}|^2 + |\overline{\lambda}|^2}{8\pi}\, 
m_{\Sigma} 
\end{equation}
while $\Gamma\left(\Sigma_f 
\rightarrow f DM\right)$ and $Br\left(\Sigma_f 
\rightarrow f DM\right)$ are the decay rate and branching 
fraction of the field $\Sigma_f$ into DM. Neglecting the final state masses, we have then: 
\begin{equation}
\label{eq:rate2}
\Gamma\left(\Sigma_f \rightarrow f DM\right)=\frac{|\lambda_{\psi\,f}|^2}{8\pi}\, m_{\Sigma},\quad
Br\left(\Sigma_f \rightarrow f DM\right)=\frac{|\lambda_{\psi\,f}|^2}{|\lambda_{\psi\,f}|^2+
|\overline{\lambda}|^{2}}
\end{equation}
We note that the freeze-in contribution is proportional to the decay rate of $\Sigma_f$ into 
DM, which is in turn proportional to $\lambda_{\Sigma f}$, independently of the other coupling, 
while the SuperWIMP contribution depends on the branching fraction of decay of $\Sigma_f$ 
into DM as well as on its mass and charge, strongly influencing the relic density 
$\Omega_\Sigma h^2$. But in general from eq.~(\ref{Omegah2-DM2}) we see that both production 
mechanisms are inefficient if the branching fraction of $\Sigma_f $ decay into DM becomes 
too small.  On the other hand a too large DM coupling to $ \Sigma_f $ can easily cause 
overproduction. 
Imposing the cosmological value of the Dark Matter density for $\psi $ fixes some definite 
ranges of the couplings $\lambda,\overline{\lambda}$ or equivalently $\tau_{\Sigma}, 
Br\left(\Sigma_f \rightarrow f DM\right)$ as a function of the mass scales, $m_\psi, m_\Sigma $.

In the next sections we will investigate the reach of LHC in the detection of
the charged field $\Sigma_f$ and investigate which signal can be expected in the parameter
regions favored by a successful cosmological DM production and possible DM decay.

\section{Collider analysis} 

Contrary to the DM, the scalar field $\Sigma_f$ is charged under SM gauge interactions 
which may give its efficient production at the LHC, if kinematically allowed. 
Since the Yukawa couplings with the quarks are much smaller than any of the 
gauge couplings, the main production channels at a proton-proton collider are 
gluon fusion into a scalar-antiscalar pair, for colored $\Sigma_f$, or 
Drell-Yan production, for the electroweakly or electromagnetically charged case.
In either cases the production rate is practically independent on the details of 
the DM model, and given just by the mass and charge of the field $\Sigma_f$.
We will estimate here the NLO production rates by computing the LO cross-section
with the package MadGraph 5~\cite{Alwall:2011uj} and correcting with a constant
NLO k-factor, depending on the channel.

For any given assignment of its quantum numbers the 
scalar particle features two kind of decay channels after its production (We are implicitly 
assuming that in the case of the $SU(2)$ doublets $\Sigma_q$ and $\Sigma_\ell$ the two 
components are mass degenerate. If this is not the case, in addition to the processes 
described below, the decay of the heavier component of the doublet 
into a W, either on or off shell, and the lighter one is open. We will better clarify 
this point later in the text). We have first of all decays into a DM particle and a 
standard model fermion, with rate proportional to $\lambda_{\psi f}^2$, which can be 
classified as follows:
\begin{eqnarray}
\Sigma_f \rightarrow & & u \psi  \quad \mbox{for} \quad \Sigma_q, \Sigma_u, \Sigma_d  \nonumber\\
\Sigma_f \rightarrow & & d \psi  \quad \mbox{for} \quad \Sigma_q, \Sigma_d  \nonumber\\
\Sigma_f \rightarrow & & l \psi  \quad \mbox{for} \quad \Sigma_\ell, \Sigma_e  \nonumber\\
\Sigma_f \rightarrow & & \nu \psi \quad \mbox{for} \quad \Sigma_\ell
\end{eqnarray}
where $l$ is a charged lepton. We see that only in the case of $ \Sigma_\ell $ the decay 
can be into  an invisible final state $ \nu\psi $, but in that case also the visible channel 
into a charged lepton is present. So the decays in general give rise to a kink in the observable 
charged track/jet due to the $\Sigma_f $ decay. 
  
The scalar field can decay as well into two SM fermions, with a rate governed by 
$\overline{\lambda}^2$, according the following channels:
\begin{eqnarray}
\Sigma_f \rightarrow & & q \overline{q}^{'}  \quad \mbox{for} \quad \Sigma_u, \Sigma_d, 
\Sigma_\ell  \nonumber\\
\Sigma_f \rightarrow & & q l  \quad \mbox{for} \quad \Sigma_q, \Sigma_d  \nonumber\\
\Sigma_f \rightarrow & & q \nu  \quad \mbox{for} \quad \Sigma_q, \Sigma_d   \nonumber\\
\Sigma_f \rightarrow & & l \overline{l} \quad \mbox{for} \quad \Sigma_\ell \nonumber\\
\Sigma_f \rightarrow & & l \nu \quad \mbox{for} \quad \Sigma_\ell, \Sigma_e 
\end{eqnarray} 
where $l$ is, again, a charged lepton while $q$ is an up or down-type quark. 
In view of the dependence of the decay rate of $\Sigma_f$ on the $\lambda_{\psi f},
\overline{\lambda}$-type coupling, a tight relation exists between possible signals at LHC 
of such decays and the constraints from the DM phenomenology, being governed by the same 
couplings. As already argued in~\cite{Arcadi:2013aba} the constraints from ID and from 
the cosmological abundance of the DM require very low values of the couplings, 
namely $\lambda, \overline{\lambda} \lesssim 
10^{-(7 \div 8)}$, thus implying that the decay vertices result displaced with 
respect to the production ones and may even lie outside the detector.

In order to determine the LHC capability of detecting this kind of decays we 
will adopt the method introduced in~\cite{Covi:2014fba}. We have then identified 
three possible detection regions, referring for definiteness to the design of the 
CMS detector (see~\cite{Chatrchyan:2008aa} for a detailed description of the 
detector), namely the Pixel and the Tracker (inner detector), and an ``outside'' region. 
The inner detector regions are sensitive to the shorter lifetimes and allow for the 
detection of displaced vertices (see for example \cite{CMS-PAS-EXO-12-037}, as well 
as~\cite{ATLAS-CONF-2013-092} for similar searches from the ATLAS collaboration). 
The ``outside'' region is instead suitable for the detection of very long-lived particles
, whose corresponding signal is constituted by outgoing tracks, thus related to the 
scalar field itself rather than its decays. On recent times searches of decay of stopped 
particles~\cite{Asai:2009ka} in the detector have been considered as well. This is an 
very intriguing possibility since this last kind of searches result complementary to the 
ones of outgoing tracks. We will further comment on this later in the text. 

We have generated several 
samples of pair produced $\Sigma_f$, at $14\,\mbox{TeV}$ of centre of mass energy, 
corresponding to different assignments of its quantum numbers and different masses, 
and determined the spatial distribution of the decay vertices from the kinematic variables 
of the events and the decay rate $\Gamma_\Sigma $ as described in~\cite{Covi:2014fba}. 
Note that we consider here a straight-line motion of the particle after production, neglecting 
the magnetic field deflection and the interactions with the intervening matter, which could 
even increase the number of decays in the inner part of the detector by bending the trajectory 
or slowing down the decaying particle~\footnote{Such effects could be captured 
only by a full detector simulation, which is beyond the scope of this paper.}.
Assuming 100 \% detector efficiency, in order to claim the discovery for a given scenario we 
have required the presence of at least 10 decay events in one of the components in which 
the detector is schematized, i.e. in the pixel or the tracker, or outside the 
detector~\footnote{Notice that this is a rather conservative requirement 
which prevents the effect of statistical fluctuations and ensures stable numerical 
results. For a pure Poisson distribution and no background 5 such events correspond 
to a discovery at the 95\% CL.}. 
The optimal scenario, and thus main focus of our analysis, is however
a ``double'' LHC detection scenario, consisting in the contemporary detection of 
at least 10 events in one of the components of the inner detector, namely pixel or tracker, 
and 10 tracks leaving the detector. This indeed would allow for a cross-check in the 
measurement of the lifetime of the scalar particle as well as a better discrimination of 
possible backgrounds.

For each of the cases considered we have performed the analysis for three definite luminosities, 
namely $25 \;{\mbox{fb}}^{-1}$, $300 \;{\mbox{fb}}^{-1}$ and $3000\; {\mbox{fb}}^{-1}$, in order 
to determine the feasibility of a next future discovery during run II, as well as the maximal 
discovery reach considering the full LHC data set and, finally, after a high luminosity run.

Since the analysis employed is not sensitive to the particular type of decay products 
of $\Sigma_f$, as long as a vertex (or kink) can be observed and happens in the
detector, we will from now on refer to a schematic setup described by just four model 
parameters: the mass $m_{\Sigma_f}$ of the scalar field, the ratio $x=m_{\psi}/m_{\Sigma_f}$ and 
two couplings $\lambda$ and $\lambda^{'}$, referring, respectively, 
to the decay of $\Sigma_f$ to the DM or to only SM states.
We have as well considered an equivalent representation in terms of the $\Sigma_f$ lifetime and
DM branching fraction, more directly connected to the phenomenological observables at the 
LHC, and also helpful to translate the results obtained in different particle physics setups. 
In either case we will find the region of parameter space where displaced
vertices and/or metastable tracks may be seen and compare it to the cosmologically
viable parameter space.

As will be discussed later, it would be very important, in order to relate an hypothetical 
LHC signal to the DM properties, to distinguish both the decay channels of $\Sigma_f$. A 
necessary condition for the identification of a particular channel is that the product of 
the total number of events times the corresponding branching fraction is large enough, in 
one of the detector regions where it is possible to observe the decay products of the scalar 
field. A proper determination of the number of events needed would require the full detector 
simulation, accounting for the capability of reconstruction of the various decay products.  

In the following subsections we will investigate separately the scenarios of color and electroweakly 
charged scalar particle.

\subsection{Colored scalar}

The first case that we are going to consider is when the field $\Sigma_f$ carries color 
charge. Colored states are expected to be more efficiently produced at the LHC. For definiteness 
we will consider a $\Sigma_d$-type field in our analysis.  As already mentioned, the 
$\Sigma_{f=q,u,d}$ pairs are produced through gluon fusion and thus the production cross 
section is substantially the same for the three kind of states, apart a possible 
enhancement in the case of $\Sigma_q$ because of multiplicity. We will indeed assume, for 
this scenario, that the two components of the doublet $\Sigma_q$ are exactly degenerate in mass. 
If this is not the case the heaviest state of the doublet could decay into the lightest one and 
a W boson, if kinematically possible, or two quarks or leptons (through an off-shell W). For 
mass splittings above $ \sim 1$ GeV this decay channel has a branching fraction substantially 
equal to one and leads to prompt decays of the heavy state in case its production is accessible 
at the LHC. For sizable enough mass splittings, such that the momentum of decay products can 
survive analysis cuts (e.g. quality of signal requirements, background discrimination cuts), 
this decay can be detected and, then, the signals discussed in the following would 
result accompanied by prompt jets or leptons. On the other hand the required mass splitting 
would imply a sensitive suppression of the pair production of the heaviest states of the doublet 
and thus a small number of this kind of events. This might not be the case of the 
$\Sigma_\ell$-type particles, as will be clarified in the next subsection. 

According to the method described above we have generated samples of 
events of $\Sigma_d$ pair production normalizing the cross-section computed by 
Madgraph with a k-factor, accounting for NLO effects, which has been taken from 
the numerical package Prospino~\cite{Beenakker:1996ed}, given the similarities with the 
supersymmetric case of a stop squark. 
From the determination of the production cross-section it is possible to infer a 
general upper limit on the LHC reach at a given luminosity $\mathcal{L}$, from the 
relation $N_{\rm ev} = \sigma_{pp \rightarrow \Sigma_{d}\Sigma^{*}_{d}} \mathcal{L} 
\geq 5$ (10 events correspond to 5 pair produced $\Sigma$) where $N_{\rm ev}$ represents 
the number of produced pairs $\Sigma_d \Sigma_d^{*}$ irrespectively of the position 
of the decay vertices.    
For the luminosities considered in our analysis the LHC reach ranges from around 
$1600\,\mbox{GeV}$ at $\mathcal{L}= 25\,{\mbox{fb}}^{-1}$ to a maximal value of 
$2200\,\mbox{GeV}$ corresponding to $\mathcal{L}=3000\,{\mbox{fb}}^{-1}$.
We have then computed the spatial distribution of the decay vertices for several 
values of the mass of $\Sigma_d$, namely 800, 1600 and 2200 GeV.
Masses below 800 GeV are currently excluded, for the range of lifetimes 
relevant for our analysis, by current searches of detector stable particles~\cite{Aad:2012pra,Chatrchyan:2013oca}. 

\begin{figure}[!ht]
\begin{center}
\subfloat{\includegraphics[width=6.5cm,height=6.5cm]{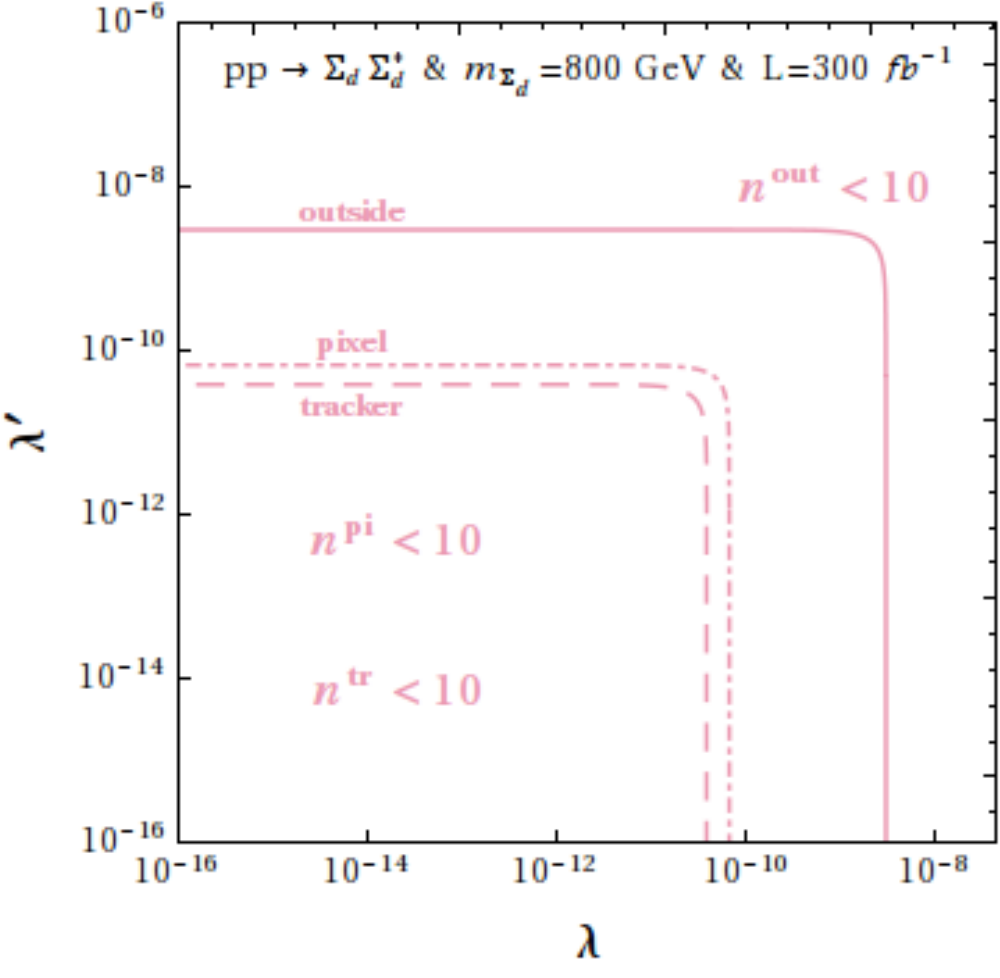}}\hspace*{1cm}
\subfloat{\includegraphics[width=6.5cm,height=6.5cm]{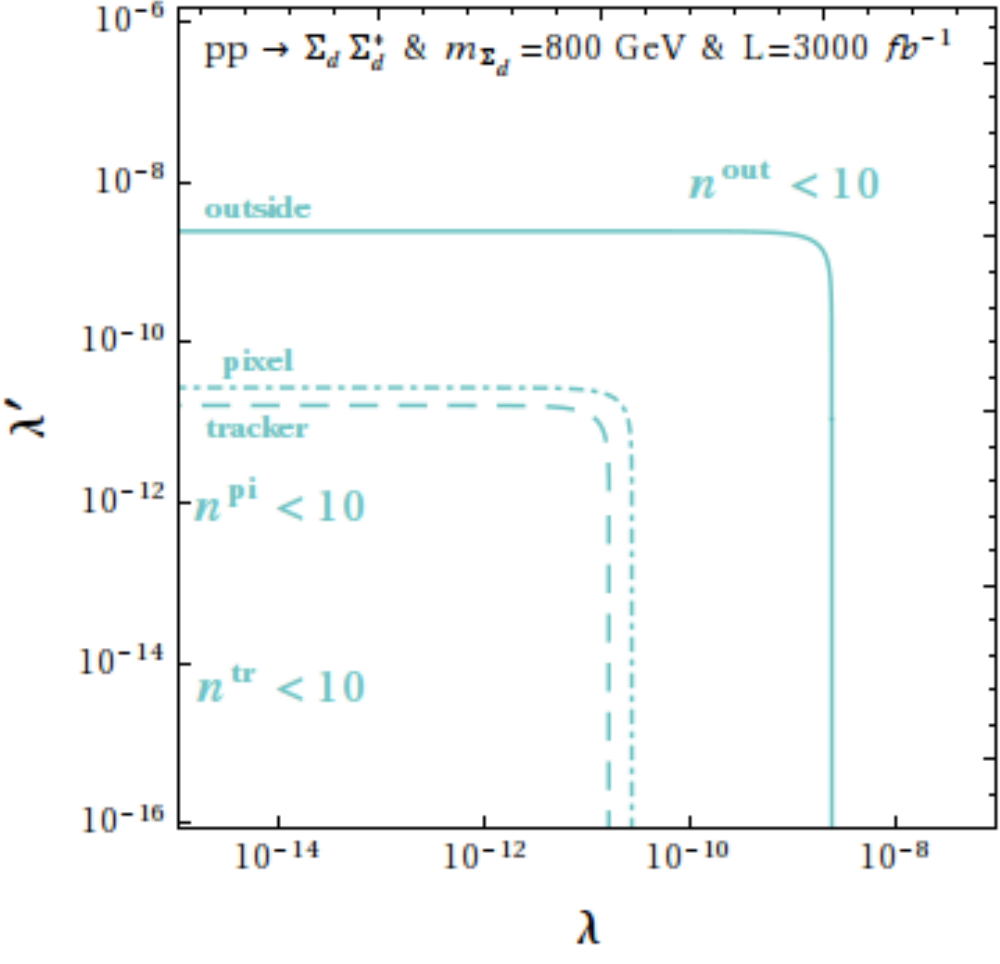}}
\caption{LHC reach in the plane of the pure SM $\lambda'$ vs DM coupling $\lambda$ for 
(from left to right) $\mathcal{L}=300,3000\,\mbox{fb}^{-1}$ for $m_{\Sigma_d}=800\,\mbox{GeV}$.
The region below the solid line corresponds to at least 10 metastable tracks, while the regions
above the dashed/dash-dotted lines to at least 10 decay events in the tracker or pixel
detector respectively.
}
\label{fig:planeLambdaVSLambdaPrimeStopm800}
\end{center}
\end{figure}

We report two examples of results of our analysis in 
fig.~(\ref{fig:planeLambdaVSLambdaPrimeStopm800}), 
for $m_{\Sigma_d} = 800$ GeV and $\mathcal{L}=300,3000 \;{\mbox{fb}}^{-1}$, 
where we have identified the region corresponding to more than 10 decay events in the 
pixel, tracker and outside the detector. We see that the contours run piecewise parallel 
to the axis, since in most of the parameter space one single coupling dominates the total 
decay rate. The detection regions for pixel and tracker are very similar, since the 
difference in volume is practically compensated by the distance from the interaction point.

The searches for displaced vertices and particles escaping from the detector are highly 
complementary: the first has a maximal reach at low lifetimes, corresponding to high 
values of $\lambda,\lambda^{'}$, while the latter is able to probe efficiently the very long 
lifetimes, i.e. low values of $\lambda,\lambda^{'}$. Combining both search strategies 
allows to cover practically the whole parameter space.

The strong requirement of detection of both types of signal, i.e. displaced
vertices in the pixel/tracker and metastable tracks, is realized only in the narrow 
regions comprised between the iso-contours in the plane ($\lambda,\lambda^{'}$) 
representing the detection of exactly 10 events in the pixel, tracker and outside 
the detector. 
The size of these strips is expected to increase with the integrated luminosity and, instead, 
to shrink once increasing the mass of the scalar, because of the lower number of particles 
produced. For the highest values of the mass, corresponding to approximately $2200 $ GeV 
the whole parameter space in the couplings might only be probed by the high luminosity 
upgrade of the LHC as long as both types of signal, either displaced vertex or 
metastable track, are considered. 

\begin{figure}[!h]
\begin{center}
\subfloat{\includegraphics[width=6.5cm,height=6.5cm]{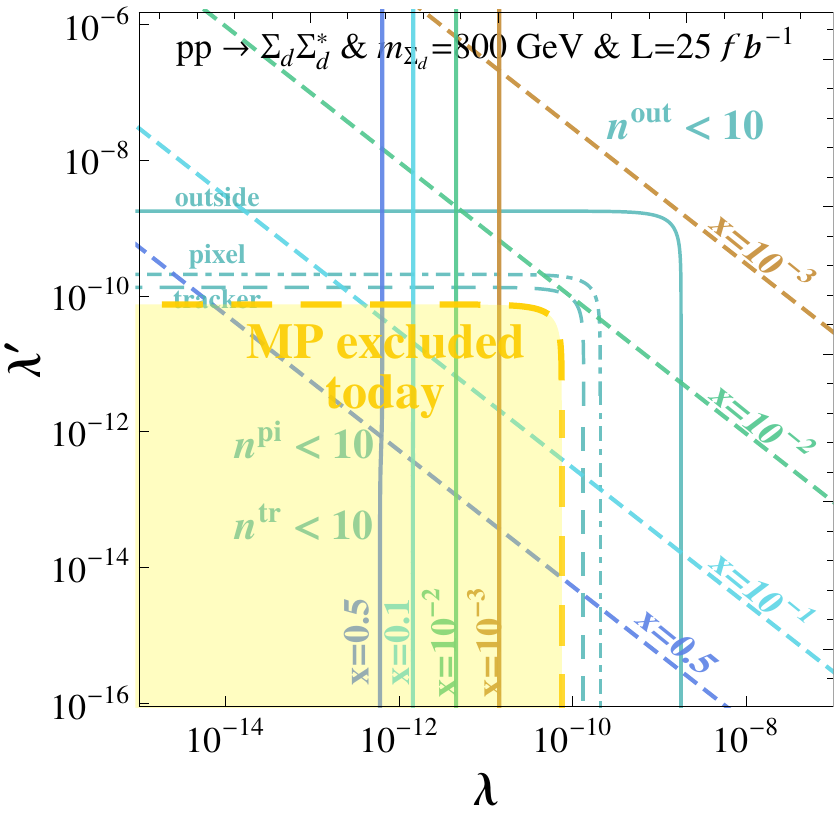}}\hspace*{1cm}
\subfloat{\includegraphics[width=6.5cm,height=6.5cm]{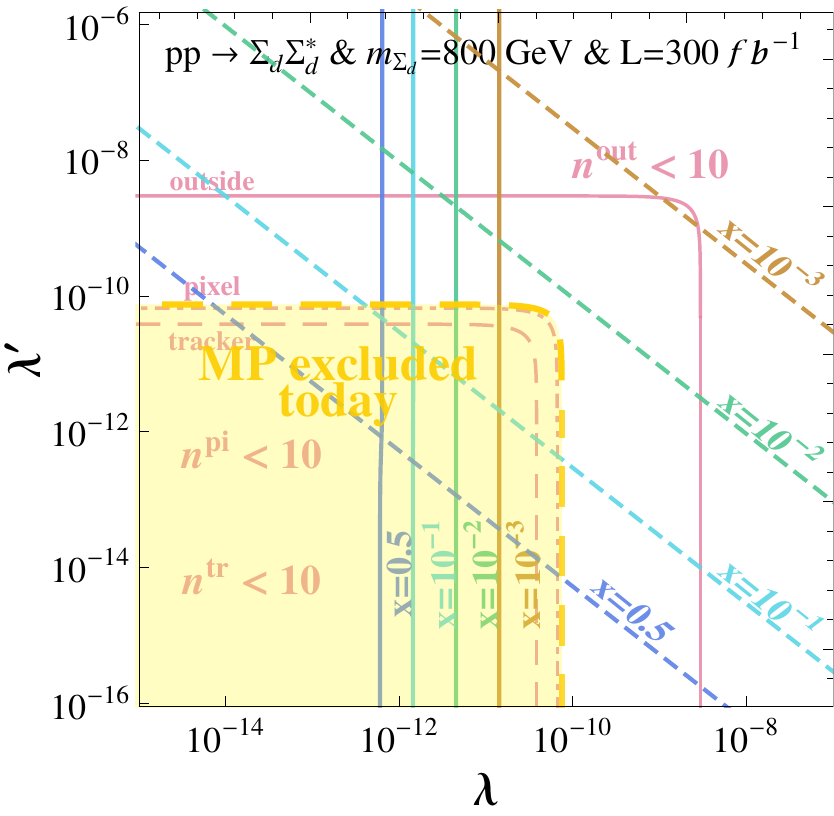}}\\
\subfloat{\includegraphics[width=6.5cm,height=6.5cm]
{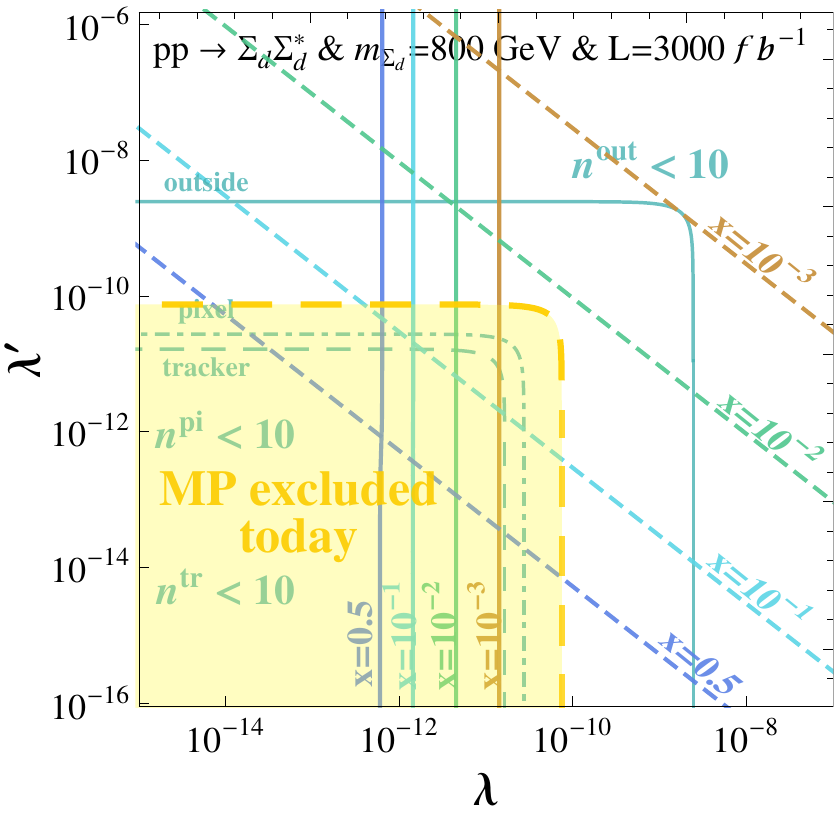}}
\caption{Contours of the correct DM relic density (solid lines) and of the reference value 
of $10^{28}\mbox{s}$ of its lifetime (dashed lines) color-coded according to the values of 
$x$ reported in the plot. A combined detection of $\Sigma_d$ at LHC and of the DM candidate 
via ID can be achieved if the lines of the DM relic density and lifetime, corresponding 
to a given value of $x$, intersect within the discovery region between the iso-contours 
labeled as pixel-tracked and outside. The plot refers to $m_{\Sigma_d}=800\,\mbox{GeV}$ and 
$\mathcal{L}=25\,{\mbox{fb}}^{-1}$ (left plot),$\mathcal{L}=300\,{\mbox{fb}}^{-1}$ (left plot), 
$\mathcal{L}=3000\,{\mbox{fb}}^{-1}$\,(bottom plot). The yellow shaded region delimited by the 
yellow thick long-dashed line is already excluded by current searches for metastable particles.}
\label{fig:800L25ALL}
\end{center}
\end{figure}

\begin{figure}[htb]
\begin{center}
\subfloat{\includegraphics[width=6.5 cm]{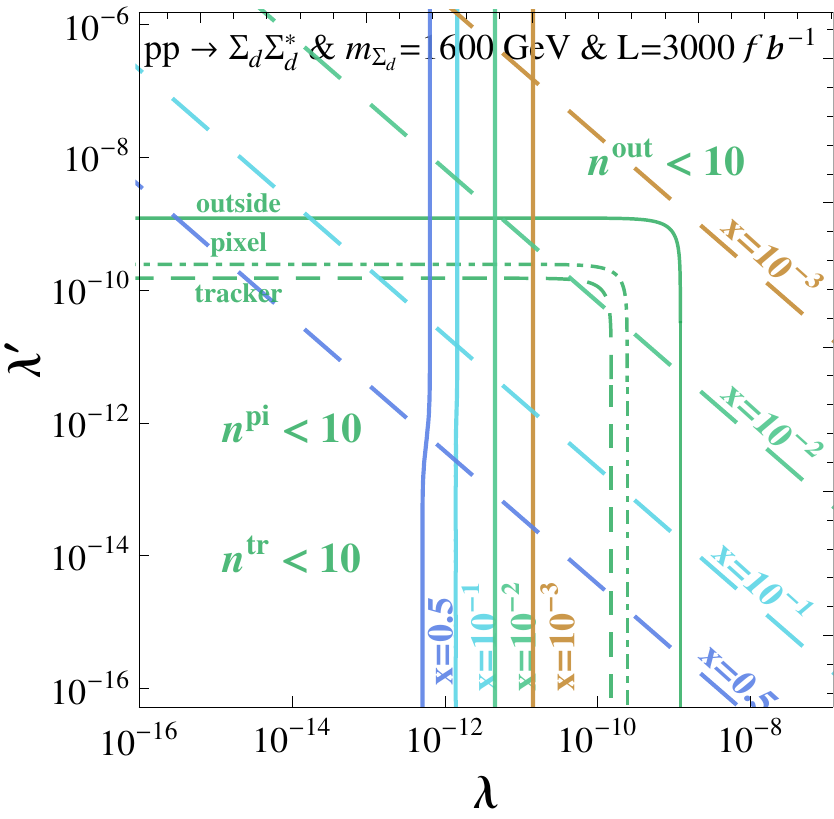}}
\hspace{10mm}
\subfloat{\includegraphics[width=6.5 cm]{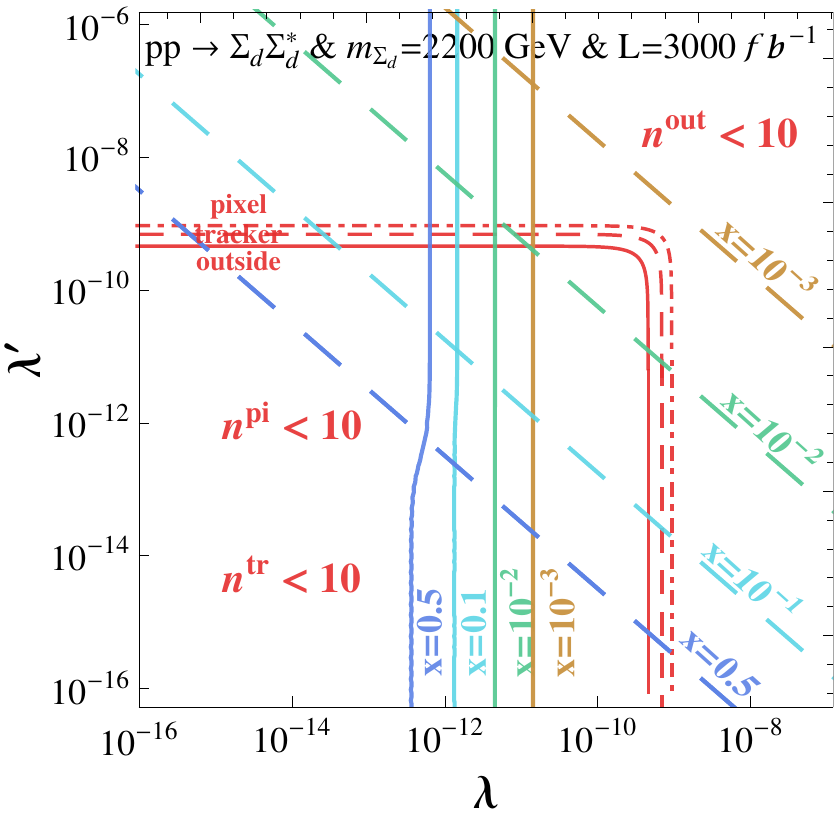}}
\caption{The same as fig.~(\ref{fig:800L25ALL}) but with $m_{\Sigma_d}=1600\,
\mbox{GeV}$ (left plot) and $m_{\Sigma_d}=2200\,\mbox{GeV}$ (right plot) and 
$\mathcal{L}=3000{\mbox{fb}}^{-1}$ in both cases.}
\label{fig:16003000ALL}
\end{center}
\end{figure}

The capability of LHC detection of a displaced decay of $\Sigma_d$
or its metastable track can be confronted with the requirement of the correct cosmological 
DM abundance via $\Sigma_d$ decay and, possibly, a detection of decaying DM.
In case of a colored scalar the correlation between the DM phenomenology and the 
LHC predictions is rather straightforward since the DM production is substantially 
dominated by the first (freeze-in) contribution in eq.~(\ref{Omegah2-DM2}) and the 
couplings can be analytically determined as function of $x$, $m_{\Sigma_d}$ and 
the DM relic density and lifetime as~\cite{Arcadi:2013aba}:
\begin{align}
\label{eq:FIMP_parameters}
& \lambda \simeq 1.59\times 10^{-12} {x}^{-1/2}{\left(\frac{g_{*}}{100}\right)}^{3/4}
{\left(\frac{\Omega_{\rm CDM}h^2}{0.11}\right)}^{1/2}g_\Sigma^{-1/2} \nonumber\\
& \nonumber\\
& \lambda^{\prime} \simeq 0.91 \times 10^{-12}\; x^{-2}
{\left(\frac{g_{*}}{100}\right)}^{-3/4}{\left(\frac{m_{\Sigma_{f}}}{1\mbox{TeV}}
\right)}^{-1/2}g_\Sigma^{1/2}{\left(\frac{\tau_\psi}{10^{27}\mbox{s}}\right)}^{-1/2}
{\left(\frac{\Omega_{\rm CDM}h^2}{0.11}\right)}^{-1/2}\;
\end{align} 
From these relations we can determine the cosmologically preferred parameter space 
in the plane of the couplings. We show indeed in fig.~(\ref{fig:800L25ALL}), as solid 
lines, the isolines of the correct value of the DM relic density for $m_{\Sigma_d}=800$ GeV 
and some values of $x$ ranging from $10^{-3}$ to $0.5$. They 
appear as vertical lines since the freeze-in mechanim is independent from $\lambda^{'}$. 
This curves can be confronted with the contours of the reach in the three detector regions. 
The panels of fig.~(\ref{fig:800L25ALL})  report the LHC reach for the three values of luminosity 
considered in our analysis. Fig.~(\ref{fig:16003000ALL}) shows an analogous analysis for the 
values $m_{\Sigma_d}=1600,2200$ GeV. In this case we have considered only 
$\mathcal{L}=3000 \;{\mbox{fb}}^{-1}$ since we expect a statistically relevant number of events 
only for this very high luminosity. 
In fig.~(\ref{fig:800L25ALL}) and~(\ref{fig:16003000ALL}) we have reported as well 
(dashed lines) the values of the DM lifetime, for the chosen combinations of parameters, 
near the present bounds. As already stated, we have assumed, for the scenario of colored scalar 
field, $u\overline{u}(d\overline{d})\nu$ as the only relevant decay channel for the DM and 
thus applied the bounds of~\cite{Garny:2012vt} in the $m_\psi>100\,\mbox{GeV}$ region and 
of~\cite{Fornengo:2013xda}~\footnote{The bounds presented here actually refer to two body 
decays in fermion pairs and then result conservative since in our setup part of the energy of 
the products is carried away by the neutrinos.} at lower masses. 
For any value of $x$, the intersection of the corresponding solid and dashed lines corresponds 
to a DM with the correct relic density and a lifetime approximately coinciding with the current 
observational bounds for the assumed dominant decay channel; consequently the parameter space 
above the DM lifetime curves may be already excluded. Notice that the actual ID exclusion 
region depends strongly on the DM decay channel and DM mass and is affected from astrophysical 
uncertainties in the propagation modeling as discussed in \cite{Garny:2012vt}.

The most favorable scenario, consisting in a multiple detection of the DM and $\Sigma_d$ decay, 
respectively in cosmic rays and at the LHC, with the latter satisfying the double LHC detection 
requirement, is potentially feasible, for a given value of the pair  $(x,m_{\Sigma_d})$, when 
the corresponding isolines of the DM relic density and lifetime intersect inside the double 
detection region. Outside this region the contemporary ID detection of DM and only one type of 
LHC signal, i.e. displaced vertices (above the white strip) or metastable tracks (below the 
white strip), is anyway still feasible. We remark, however, that the region below the white 
strip, corresponding to very long lifetimes, is already constrained, for the lower values of 
the mass of the scalar, by current searches of detector stable particles. We have reformulated, 
for the scenario under consideration, these constraints using the procedure described 
in~\cite{Covi:2014fba} and reported the excluded region in fig.~(\ref{fig:800L25ALL}). The 
region below the ``double'' detection strip is nearly ruled out for $m_{\Sigma_d}=800$ GeV. 
The limit from detector stable particles weakens very quickly with increasing mass of the 
scalar particle and it is substantially irrelevant for masses above 1 TeV.   

The double LHC detection region corresponds, for $m_{\Sigma_d}=800\,\mbox{GeV}$, to a rather 
definite range of values of $x$ comprised between $10^{-2}$ and $10^{-1}$. This range is 
reduced at higher values of $m_{\Sigma_d}$ because of the decreased size of the LHC double 
detection strip. For the highest possible value 
$m_{\Sigma_d}$ the combined detection prospects are substantially limited to $x \simeq 10^{-2}$. 

\begin{figure}[htb]
\begin{center}
\subfloat{\includegraphics[width=6.5cm]{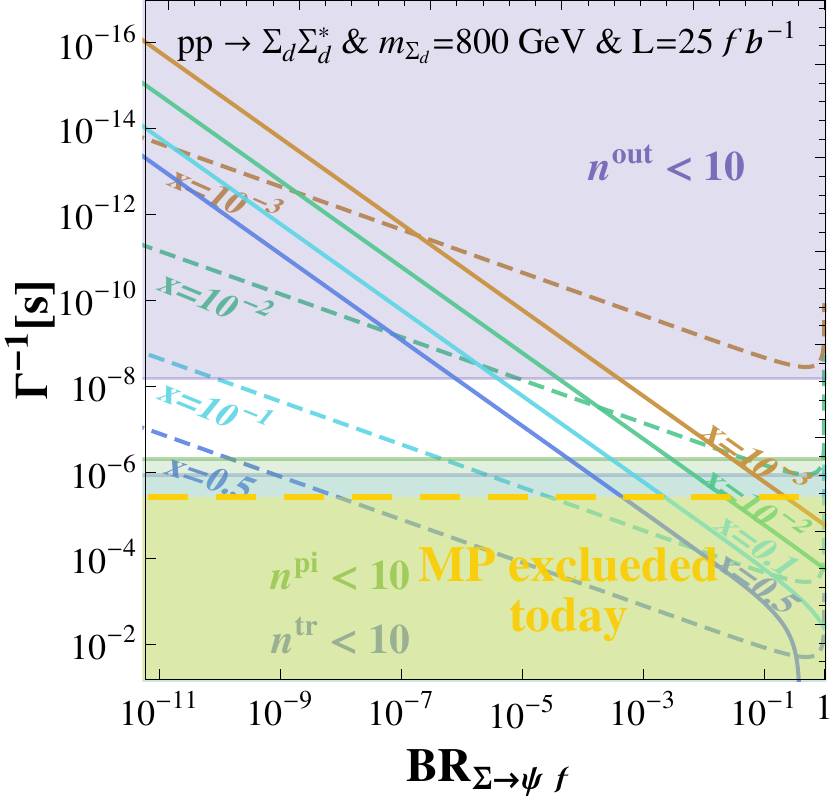}}
\hspace{10mm}
\subfloat{\includegraphics[width=6.5cm]{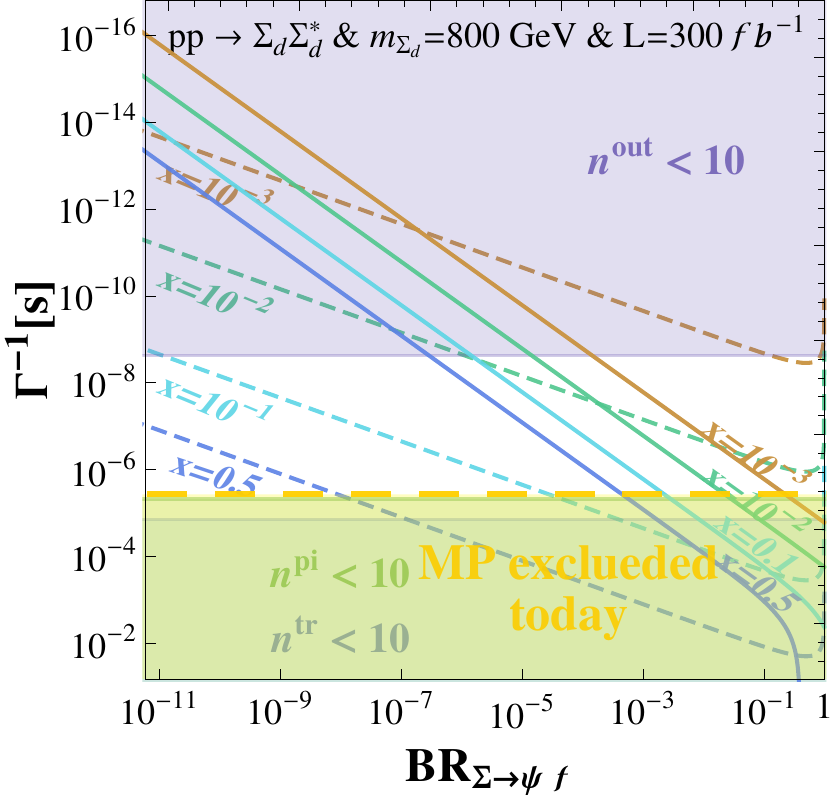}}
\caption{LHC detection reach compared with the constraints from DM phenomenology in the 
plane $\left(Br\left(\Sigma_d \rightarrow \mbox{DM}\right), \Gamma_{\Sigma_d}^{-1}\right)$. 
The shaded green (magenta) region corresponds to more than $10$ decay events happening 
in the pixel or tracker (outside the detector). The
double detection region is the white strip comprised 
between the shaded regions. We have fixed $m_{\Sigma_d}=800\,\mbox{GeV}$ and 
$\mathcal{L}=25{\mbox{fb}}^{-1}$ (left plot) and $\mathcal{L}=300\,{\mbox{fb}}^{-1}$ (right plot). 
The yellow shaded region, below the thick
long-dashed yellow line, is excluded by current bounds on metastable particles.}
\label{fig:800BRGAMMA}
\end{center}
\end{figure}

\begin{figure}[!h]
\begin{center}
\includegraphics[width=6.5cm,height=6.5cm]{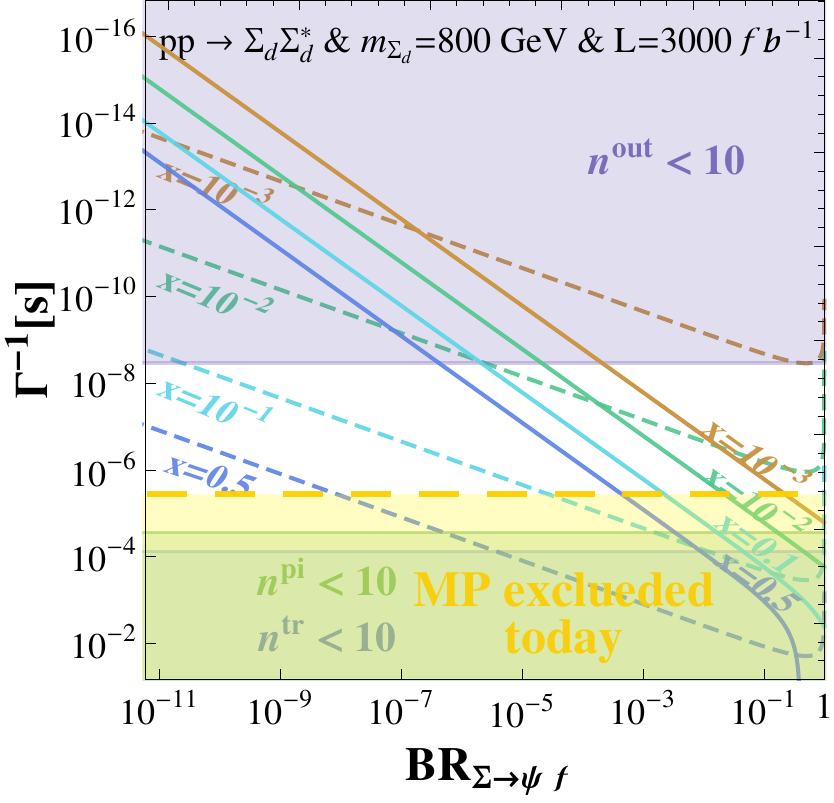}
\label{Stop800FIN}
\end{center}
\begin{center}
$\begin{array}{cc}
\includegraphics[width=6.5cm,height=6.5cm]{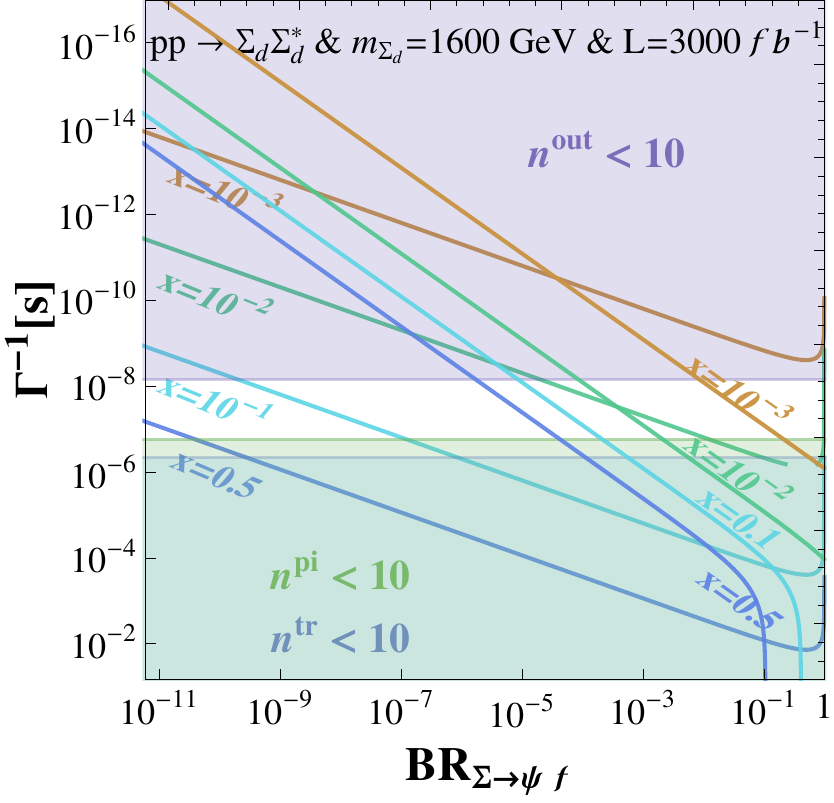}&
\hspace{10mm}\includegraphics[width=6.5cm,height=6.5cm]
{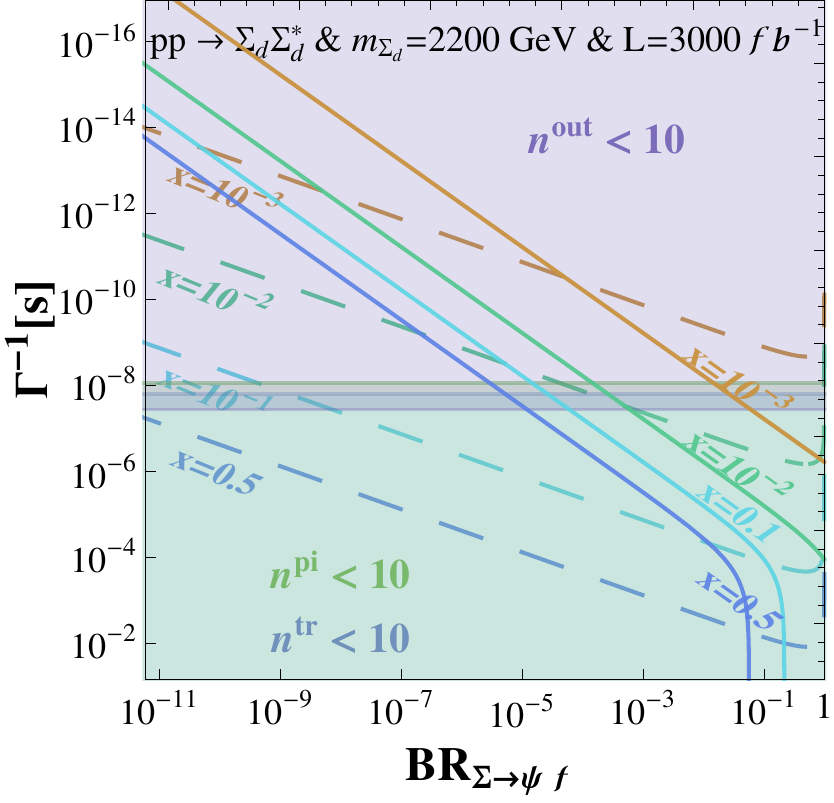}
\end{array}$
\caption{The same as fig.~(\ref{fig:800BRGAMMA}) but with $m_{\Sigma_d}=800\,
\mbox{GeV}$ (upper plot), $m_{\Sigma_d}=1600\,
\mbox{GeV}$ (left plot) and $m_{\Sigma_d}=2200\,\mbox{GeV}$ (right plot) 
and $\mathcal{L}=3000{\mbox{fb}}^{-1}$ in all cases.}
\label{fig:16003000BRGAMMA}
\end{center}
\end{figure}

It is also interesting to reformulate the previous results in a more model independent way 
in terms of the pair $\left(\Gamma_{\Sigma}^{-1}(\mbox{s}),Br\left(\Sigma_d \rightarrow \mbox{DM}
\right)\right)$ as done in fig.~(\ref{fig:800BRGAMMA}), (\ref{fig:16003000BRGAMMA}).
Using these parameters the LHC detection regions are just delimited by horizontal
lines of constant $\Gamma_{\Sigma}$ values.
The green (violet) shaded regions in the plot represent the regions in which it is 
possible to detect more than 10 events in the pixel/tracker (outside region). 
``Double'' signals are accessible instead in the middle white strip.
The combined LHC detection of the $\Sigma_d$, in at least one of the two channels,
and DM indirect
detection are again achieved whenever the isolines (dash-dotted) of the reference DM 
lifetime and of the correct relic density (solid) for a fixed value of $x$ cross in 
the LHC ``double'' discovery region.
Above this strip it is still possible to observe displaced vertices at the
LHC and have an ID DM signal for small values of $x$. At large values of $x$, instead, 
only metastable particle signals are compatible with DM ID.

As evident from the plots, simultaneous ID/LHC detection, with the latter in the form of 
a double signal, i.e. displaced decay plus metastable track, could be achieved in 
the next future, namely for luminosities up to $300{\mbox{fb}}^{-1}$, for values 
of $Br_{\Sigma_d \rightarrow d \psi}$ lower than $10^{-3}$. As a consequence the 
observation at the LHC of both the DM and only SM decay channels of $\Sigma_d$ 
appears difficult but it is not a priori excluded provided that there is a 
high statistics.

In order to investigate this possibility we have performed a more focused study on 
a benchmark set of parameters, namely $m_{\Sigma_d}=800\,\mbox{GeV}$, corresponding to the 
maximal production cross-section, $m_{\psi}=10\,\mbox{GeV}$ and $\lambda$ and $\lambda^{'}$
fixed to reproduce the correct DM relic density and a DM lifetime of approximately 
$10^{28}$ seconds. We have reported on table~(\ref{tab:colored_benchmark}) the number of 
decay events, expected at the three values of the luminosity, namely 25, 300 and 3000 
${\mbox{fb}}^{-1}$, together with an estimate of the number of events in the two different 
type of decay channels, namely in DM plus SM fermion and two SM fermions. 

\begin{table}
\begin{center}
\begin{tabular}{|c|c|c|c|}
\hline
Part of detector & Total & $\Sigma \rightarrow DM$ & $\Sigma \rightarrow \mbox{SM only}$ \\
\hline
$\mathcal{L}=25 {\mbox{fb}}^{-1}$\\
\hline
Pixel & 31 & 0 & 31 \\
\hline
Tracker & 63 & 0 & 63 \\
\hline
Out & 454 & 1 & 453 \\
\hline
$\mathcal{L}=300 {\mbox{fb}}^{-1}$\\
\hline
Pixel & 378 & 0 & 378 \\
\hline
Tracker & 752 & 1 & 751 \\
\hline
Out & 5445 & 6 & 5439 \\
\hline
$\mathcal{L}=3000 {\mbox{fb}}^{-1}$\\
\hline
Pixel & 3785 &  4 & 3781 \\
\hline
Tracker & 7522 & 8 & 7514 \\
\hline
Out & 54446 &  57 & 54389 \\
\hline
\end{tabular}
\caption{Number of decay events, total as well as separately in the two kind of decay channels 
(DM or SM only), which is expected to be observed, in the three detection regions, at the 
indicated values of the luminosity, for a $\Sigma_d$ production scenario corresponding to the 
benchmark set of masses $m_{\Sigma_d}=800$ GeV and $m_\psi=10$ GeV. The coupling $\lambda$ and 
$\lambda^{'}$ have been chosen in such a way that the DM achieves the correct relic density, 
through the freeze-in mechanism, and its lifetime is $10^{28}$ s, just beyond present ID limits. 
The values of $\lambda$ and $\lambda^{'}$ are, respectively, $1.8\times 10^{-11}$ and $5.5
\times 10^{-10}$.}
\label{tab:colored_benchmark}
\end{center}
\end{table}

\begin{table}
\begin{center}
\begin{tabular}{|c|c|c|c|}
\hline
Part of detector & Total & $\Sigma \rightarrow DM$ & $\Sigma \rightarrow \mbox{SM only}$ \\
\hline
$\mathcal{L}=25 {\mbox{fb}}^{-1}$\\
\hline
Pixel & 0 & 0 & 0 \\
\hline
Tracker & 0 & 0 & 0 \\
\hline
Out & 178 & 19 & 159 \\
\hline
$\mathcal{L}=300 {\mbox{fb}}^{-1}$\\
\hline
Pixel & 0 & 0 & 0 \\
\hline
Tracker & 0 & 0 & 0 \\
\hline
Out & 2133 &  222 & 1911 \\
\hline
$\mathcal{L}=3000 {\mbox{fb}}^{-1}$\\
\hline
Pixel & 0 &  0 & 0 \\
\hline
Tracker & 0 & 0 & 0 \\
\hline
Out & 21334 &  2225 & 19109 \\
\hline
\end{tabular}     
\caption{The same as tab.~(\ref{tab:colored_benchmark}) but for $m_\Sigma=1$ TeV, $x=0.5$ and 
the couplings $\lambda$ and $\lambda^{'}$ set to, respectively, $1.2 \times 10^{-12}$ and 
$3.6 \times 10^{-12}$, in order to achieve the correct DM relic density and a DM lifetime of 
$10^{28}$ s.} 
\label{tab:colored_benchmark_1}
\end{center}
\end{table}

As evident, no events of decay into DM can be observed inside the detector (namely pixel or 
tracker), even in a high luminosity run of LHC. This is a consequence of the very 
suppressed branching ratio of decay of $\Sigma_d$ into DM, $\approx 10^{-3}$, required 
to reconcile the correct DM relic density with observable decays of the latter.  
A potential observation of decay of $\Sigma_d$ inside the detector seems thus limited 
to the pure SM channel. Possible dedicated searches should be then optimized for the search 
of displaced vertices with multi-jet/leptons and low amounts of missing energy. 

The observation of the DM channel might be feasible considering the possibility of observing 
decays of stopped particles in the detector. Indeed we have until now assumed that the decays 
of $\Sigma_d$ can be observed in the pixel/tracker region while beyond those only escaping 
tracks associated to the particle $\Sigma_d$ itself are detected. In reality 
colored/electromagnetically charged metastable particles lose energy by interacting with the 
detector material and a fraction of them can be stopped inside the detector itself. Their late 
time decays can be observed in the intervals between the collisions of proton 
beams~\cite{Aad:2013gva,Chatrchyan:2012dxa}.
These kind of searches are characterized by rather low efficiencies, in particular because of 
the low, namely $\lesssim 10\%$, fractions of stopped particles; as a consequence one should 
focus on the regions in which the number of decays outside the pixel/tracker regions is maximal, 
possibly renouncing to a statistically significant number of events inside the detector. So 
searches for stopped particles are in some sense complementary to the searches discussed in 
this work. At the same time, as can be noted e.g. in fig.~(\ref{fig:800BRGAMMA}), 
(\ref{fig:16003000BRGAMMA}), in the regions where the number of particles 
decaying outside the detector is maximal, the isolines of DM lifetime and relic density cross at 
values of the branching ratios of order 0.5 and thus a contemporary detection of both decays 
of $\Sigma_d$ is feasible provided the presence of statistically relevant population of 
stopped particles.

This possibility is confirmed by the outcome of the analysis reported in 
tab.~(\ref{tab:colored_benchmark_1}), where a benchmark with $(m_\psi,m_{\Sigma_d})=(500,1000)$ 
GeV has been considered and the couplings set in order to achieve, besides the correct DM 
relic density, a lifetime of the DM of $10^{28}$ s and a branching ratio of decay of 
$\Sigma$ into DM of approximately $10\%$. For luminosities above $300 \;{\mbox{fb}}^{-1}$, 
a sizable number of events could be observed in both the decay channels of $\Sigma_d$, 
possibly compensating the low efficiency in the detection of stopped particles. In order to 
quantitatively explore this scenario an analysis accounting for the typology of  decay products 
of $\Sigma_d$ as well as a simulation of the detector are however needed. 
This is beyond the scope of this work and will be left to a future study. 

The contemporary detection of both decay channels of $\Sigma_d$ can be feasible as well 
at very low values of $x$, thus corresponding to very light DM particles. Indeed in such a 
case it is possible to set the two couplings $\lambda$ and $\lambda^{'}$ to comparable values 
without conflicting with ID constraints because of the strong enhancement, namely 
$\propto x^{-5}$ of the DM lifetime. We also notice that higher values of the coupling 
$\lambda$ are favored by the DM relic density since it tends to be suppressed as $x$. 
In such scenario we thus expect $\Sigma_f$ particles to be relatively short-lived and thus 
most of the decay events happen in the pixel/tracker region. 
This is confirmed by the results shown in tab.~(\ref{tab:colored_benchmark_2}) in which we 
have considered a benchmark model with the very low value $x=10^{-6}$. As we expected, most 
of the decay events lie in the inner detector with statistically relevant populations in 
both the decay channels. For luminosities above $300{\mbox{fb}}^{-1}$ it is nonetheless possible 
to observe more than 10 escaping tracks; the LHC thus provides an optimal reconstruction of 
the properties of the scalar field. On the other hand for the very low value of $x$ considered the 
Indirect Detection of DM is not possible since its lifetime largely exceeds present and next 
future experimental sensitivity. 
As a consequence, in this kind of setup, LHC is the only probe of the model under consideration.

\begin{table}
\begin{center}
\begin{tabular}{|c|c|c|c|}
\hline
Part of detector & Total & $\Sigma \rightarrow DM$ & $\Sigma \rightarrow \mbox{SM only}$ \\
\hline
$\mathcal{L}=25 {\mbox{fb}}^{-1}$\\
\hline
Pixel &  49 & 8 &  41 \\
\hline
Tracker & 58 & 9 & 49 \\
\hline
Out & 2 & 0 & 2 \\
\hline
$\mathcal{L}=300 {\mbox{fb}}^{-1}$\\
\hline
Pixel & 585 & 92 & 493 \\
\hline
Tracker & 692 & 109 & 583 \\
\hline
Out &  25 & 4 & 21 \\
\hline
$\mathcal{L}=3000 {\mbox{fb}}^{-1}$\\
\hline
Pixel & 5848 & 925 & 4923 \\
\hline
Tracker & 6923 &  1094 & 5829 \\
\hline
Out & 250 & 40 & 210 \\
\hline
\end{tabular}
\caption{The same as tab.~(\ref{tab:colored_benchmark}) but for $m_\Sigma=1$TeV, $x=10^{-6}$ 
and the couplings $\lambda$ and $\lambda^{'}$ set to, respectively, $6.5 \times 10^{-10}$ and 
$1.5 \times 10^{-9}$. For this choice of parameters $Br(\Sigma_d \rightarrow DM)\sim 0.15$ and 
the DM relic density is entirely achieved through the freeze-in mechanism. The DM lifetime 
instead exceeds of many orders of magnitude the sensitivity of present and next future 
detectors.}
\label{tab:colored_benchmark_2}
\end{center}
\end{table}

\subsection{EW-charged scalar}

In this section we present a similar analysis for $\Sigma_{e, \ell}$-type scalar field. 
We will first assume, analogously to the previous scenario, that the two components of 
the $SU(2)_L$ doublet are exactly degenerate in mass, such that the analysis can be 
carried out with the exact same steps as before.

The case in which, instead, a sizable mass splitting is present is also phenomenologically 
intriguing. Indeed one of the components of $\Sigma_\ell$ is electrically neutral and 
remains undetectable at LHC in case of decays outside the Pixel/Tracker region, thus behaving 
like an additional DM component. Additional interesting collider signatures can arise in the 
case in which both the components are accessible to LHC production and decays in W boson 
(either on- or off-shell) are allowed. We will comment on this possibility at the end of 
this subsection.  

\begin{figure}[ht!]
\begin{center}
\includegraphics[width=7.0 cm]{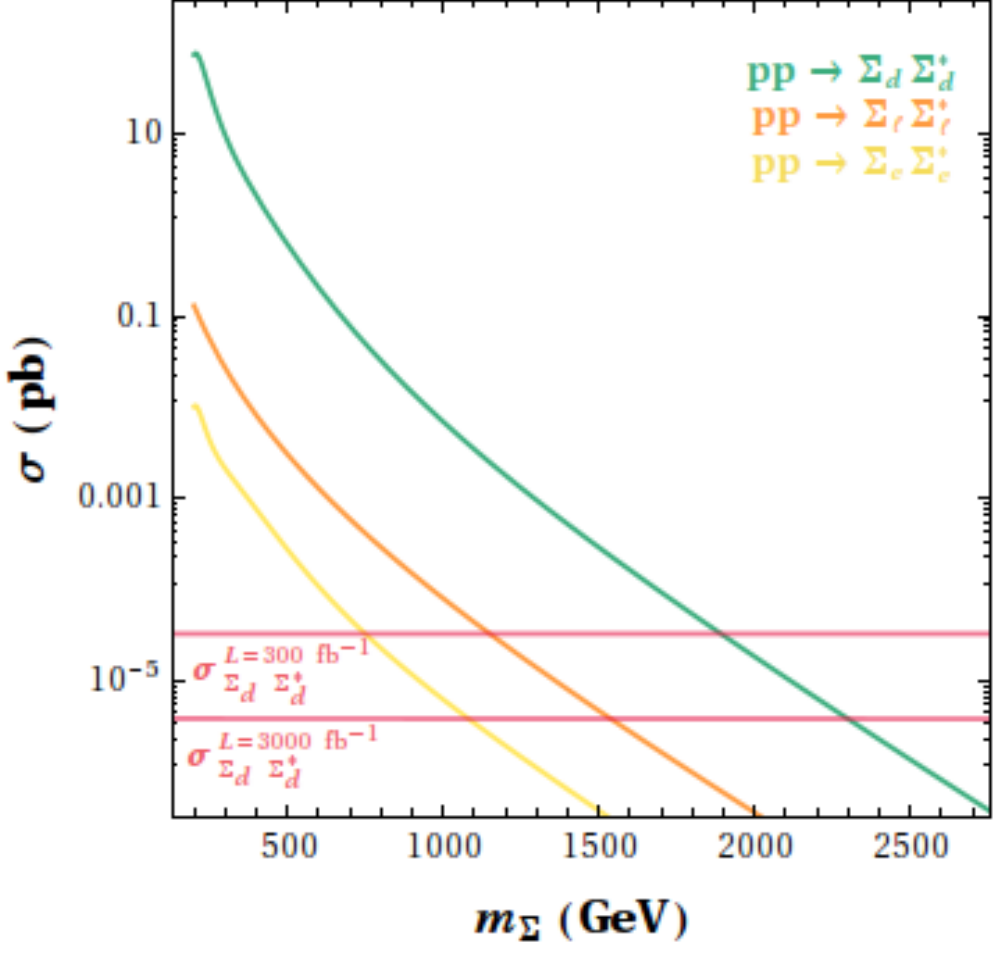}
\vspace{-3mm}
\caption{NLO cross-section for several kinds of $\Sigma_f$ field, namely $\Sigma_d$, $\Sigma_\ell$ and $\Sigma_e$. 
We report as well the minimal value of the production cross-section, for two values of the luminosity, 
namely 300 and 3000 ${\mbox{fb}}^{-1}$, needed to give 5 pairs of particles.}
\label{fig:CrossSectComparison}
\end{center}
\end{figure}

We show in fig.~(\ref{fig:CrossSectComparison}) the production cross-section of the 
$\Sigma_\ell $ and $\Sigma_e$ type fields, compared with the one of $\Sigma_d$. As we can 
see the production cross-sections are sensitively lower, with respect to the colored case, with a 
maximal mass reach, corresponding to the high luminosity upgrade of LHC, of approximately 
$1400 \,\mbox{GeV}$.  
At the same time the limits from detector stable particles are substantially relaxed with respect 
to the case of colored particles so that we can consider values of the mass as low as 
$300-400 \,\mbox{GeV}$.

We also notice that the production cross-section of $\Sigma_\ell$ pairs is sensitively larger 
with respect to the case of a $SU(2)_L$ singlet. This enhancement is substantially due to the 
process $pp \rightarrow \Sigma_\ell^{\pm}\Sigma_\ell^{0}$, with  $\Sigma^{\pm}_\ell$ and 
$\Sigma_\ell^{0}$ being, respectively, the electrically charged and neutral components 
of the $SU(2)$ doublet. On the other hand the increase in the expected number of events 
depends on the lifetime of $\Sigma_\ell$. It is indeed maximal at shorter lifetimes, when 
the scalar field decays prevalently in the inner detector, since displaced vertices can be 
detected both for electrically charged and neutral mother states, while it is more moderate at 
the longest lifetimes since only the charged component of the $SU(2)_L$ doublet can manifest as 
metastable tracks while the neutral one escapes detection.    

\begin{figure}[htb]
\begin{center}
\subfloat{\includegraphics[width=6.5cm]{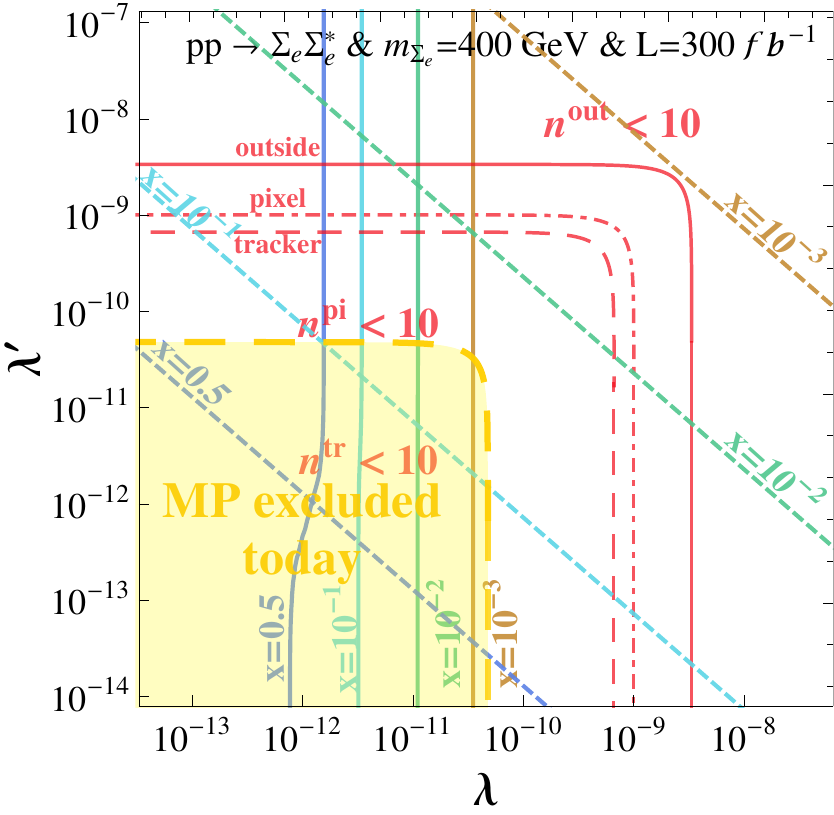}}
\hspace{10mm}
\subfloat{\includegraphics[width=6.5cm]{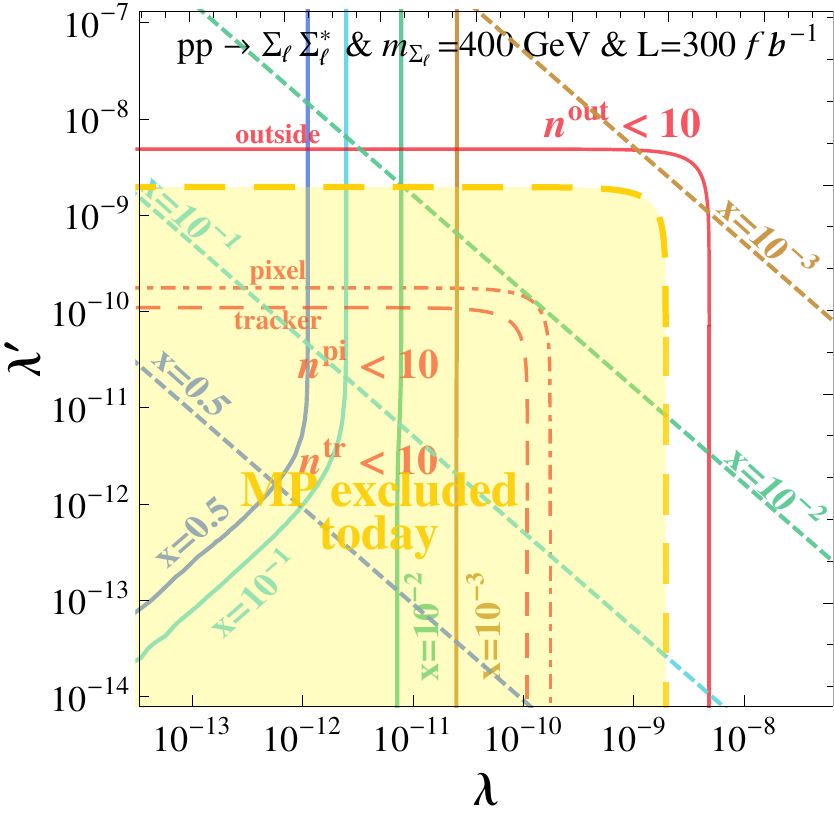}}
\caption{LHC reach, in the plane $(\lambda, \lambda^{'})$, in the Pixel (red dot-dashed lines), 
Tracker (red dashed lines) and outside the detector (red solid lines), for an integrated 
lumino\-sity of $300{\mbox{fb}}^{-1}$ for a $\Sigma_e$ (left panel) and a $\Sigma_\ell$ 
(right panel) scalar field of mass 400~GeV. The solid lines represent the cosmological value of 
the DM relic density for $x=\left\{10^{-3}, 10^{-2}, 0.1, 0.5\right\}$, while the short-dashed 
lines represent a reference value of $10^{28}\mbox{s}$, approximately correspon\-ding to the 
current experimental sensitivity on the DM lifetime for the considered set of values of  $x$. 
The yellow region below the thick long-dashed yellow line is excluded by searches of metastable 
particles.}
\label{fig:SigmaEW400}
\end{center}
\end{figure}

\begin{figure}[htb]
\begin{center}
\subfloat{\includegraphics[width=6.5cm]{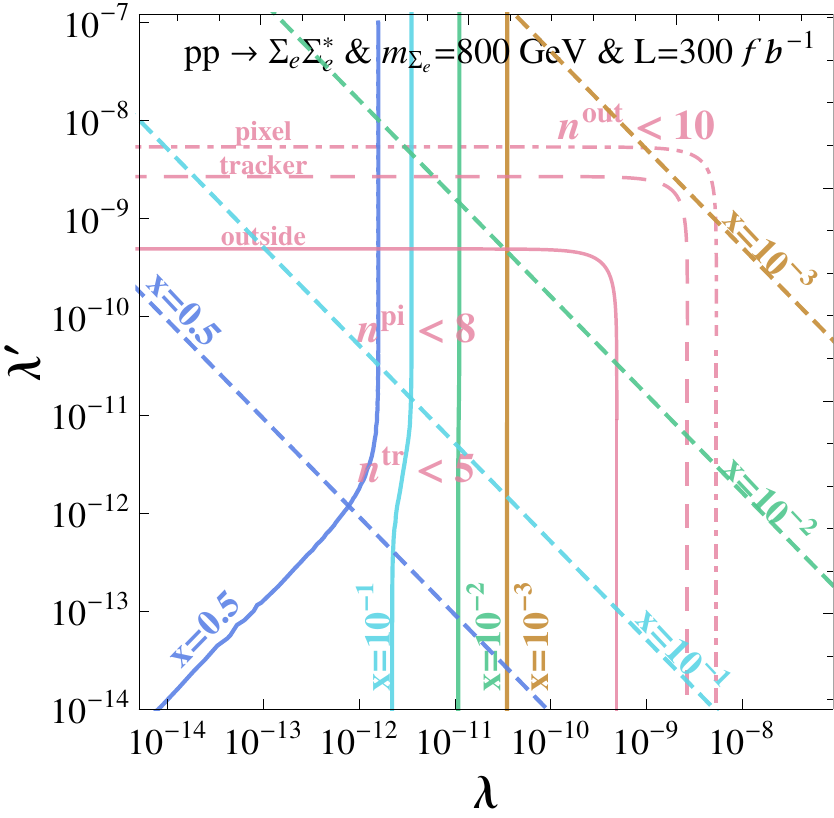}}
\hspace{10mm}
\subfloat{\includegraphics[width=6.5cm]{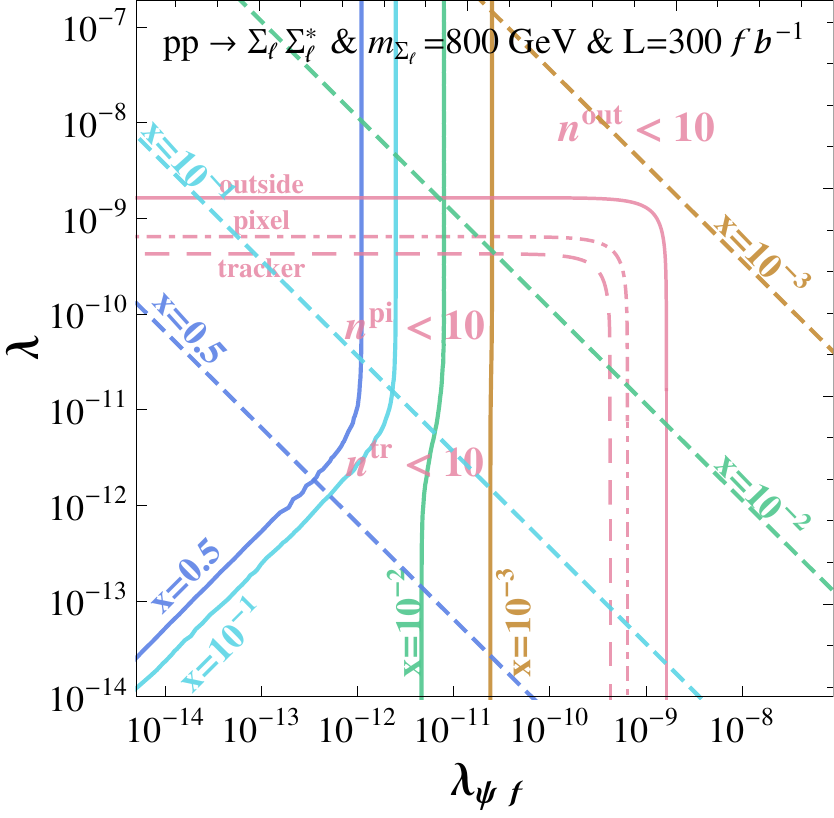}}
\caption{The same as fig.~(\ref{fig:SigmaEW400}) but for $m_{\Sigma_{e,\ell}}=800$ GeV.}
\label{fig:SigmaEW800}
\end{center}
\end{figure}

Fig.~(\ref{fig:SigmaEW400}) and~(\ref{fig:SigmaEW800}) show the LHC reach, in the three detection 
regions, as function of the couplings $\lambda$ and $\lambda^{'}$ and for some fixed values of 
$x$, for two values of $m_{\Sigma_{\ell,e}}$, namely 400 and 800 GeV. In the lower mass scenario 
it is again present an upper bound on the lifetime of the scalar field coming from current searches 
of disappearing tracks. Notice that this last limit is stronger in the case of $\Sigma_\ell$ type 
field as consequence of the higher cross-section at a given value of the mass 
(see~\cite{Covi:2014fba} for details.).  

Contrary to the colored case it is possible to have sizable or even dominant contribution 
from the SuperWIMP mechanism at masses accessible to the LHC production. However, as evident 
form the plot, this occurs only at very small values of the couplings, such that the decays 
of the scalar field occur substantially only outside the detector. 

\begin{figure}[htb]
\begin{center}
\subfloat{\includegraphics[width=6.5cm]{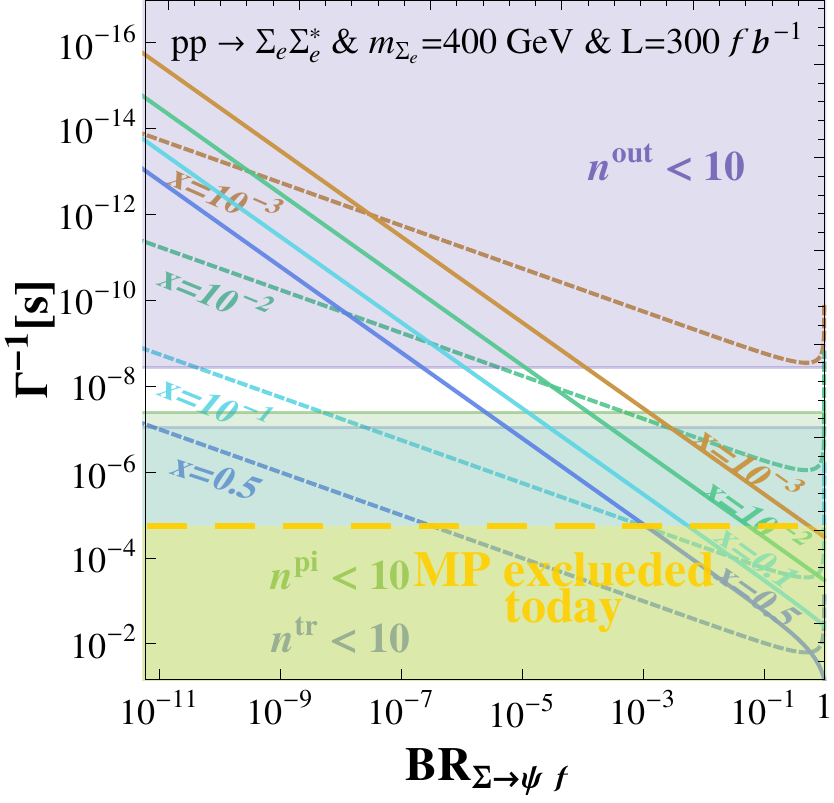}}
\hspace{10mm}
\subfloat{\includegraphics[width=6.5cm]{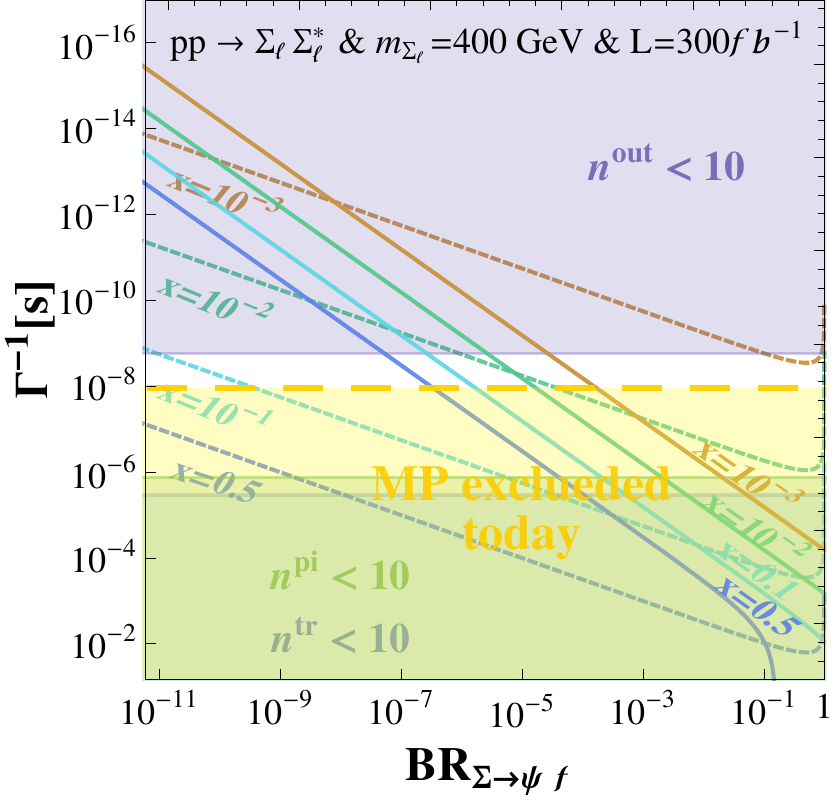}}
\caption{LHC reach for $\Sigma_e$ (left) and $\Sigma_\ell$ (right) in the plane 
$\left(Br\left(\Sigma_{e,\ell} \rightarrow \mbox{DM}\right), \Gamma_{\Sigma}^{-1}\right)$
for integrated luminosity of $300 {\mbox{fb}}^{-1}$ and 
$m_{\Sigma_{e, \ell}}=400$ GeV. The violet, green and blue region represent the regions 
where less then 10 events are expected, respectively in the 'outside', tracker and pixel 
regions. The 'double detection' region, defined in the main text, is thus the white 
strip between the shaded regions. The solid and dashed lines give
again the co\-smological relic density and the reference DM lifetime for the set 
$x=\left\{10^{-3}, 10^{-2}, 0.1, 0.5\right\}$. The yellow region below the thick dashed 
yellow line is excluded by current searches of metastable particles.} 
\label{fig:SigmaEW400-BrGamma}
\end{center}
\end{figure}

\begin{figure}[htb]
\begin{center}
\subfloat{\includegraphics[width=6.5cm]{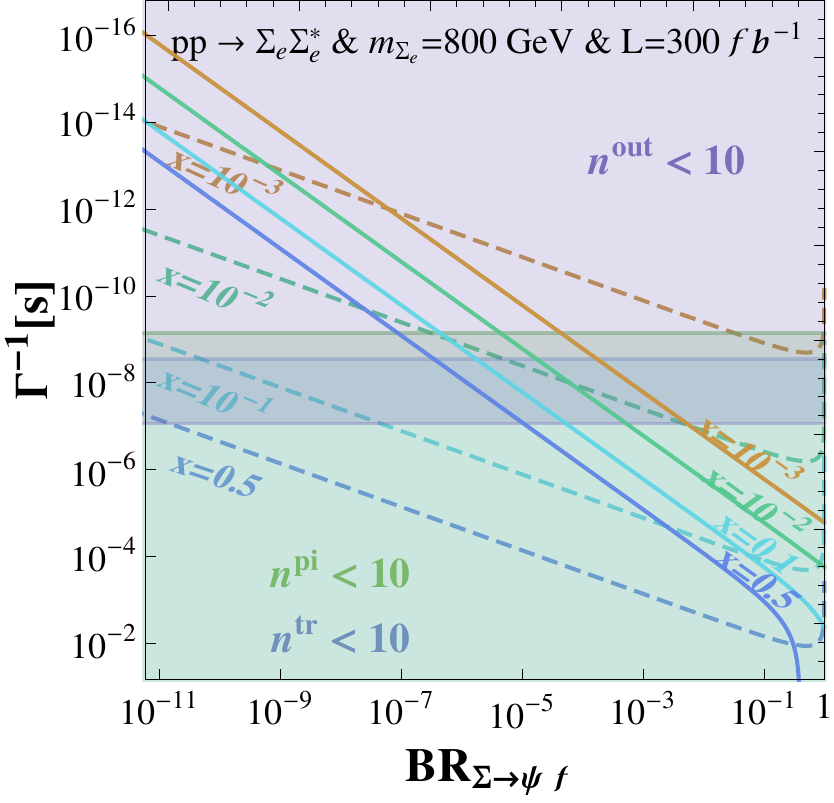}}
\hspace{10mm}
\subfloat{\includegraphics[width=6.5cm]{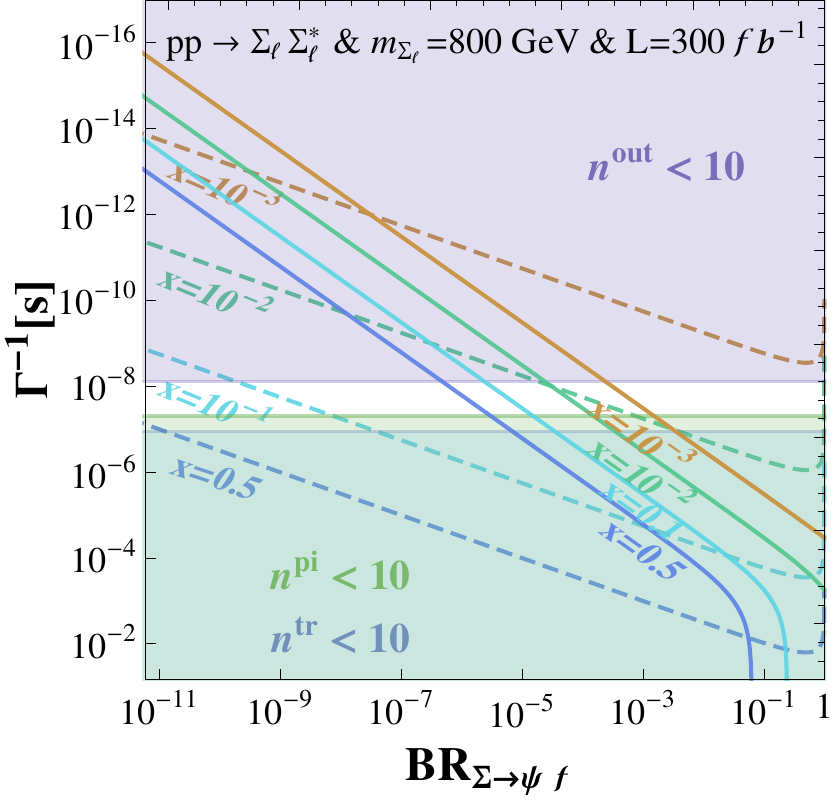}}
\caption{The same as fig.~(\ref{fig:SigmaEW400-BrGamma}) but for $m_{\Sigma_{e,\ell}}=800$ GeV. 
A 'double detection' region is present only in the case of $\Sigma_\ell$-type field.}
\label{fig:SigmaEW800-BrGamma}
\end{center}
\end{figure}

We have as well reformulated our results in the plane 
(BR($\Sigma_f \rightarrow$ DM)-$\Gamma^{-1}_\Sigma $). 
As evident from  fig.~(\ref{fig:SigmaEW400-BrGamma}) and (\ref{fig:SigmaEW800-BrGamma}) the 
LHC ``double'' detection region defined in the previous sections is extremely narrow, as 
consequence of the lower number of expected events, due to the lower production cross-sections, 
and it is already closed, for masses of 800 GeV, in the case of $\Sigma_e$-type field, which 
thus features extremely poor detection prospects. We also notice that the crossing of curves 
of the relic density and the DM lifetime occurs, in the double detection strip, at values of 
the branching ratio at most of the order of $10^{-3}$ which again makes very 
difficult (likely impossible) the detection of the decay channel of $\Sigma_{\ell,e}$ into DM.

Analogously to the previous section, we have investigated the prospects of detection of 
$\Sigma_{\ell,e}$ from a more quantitative perspective focussing on a benchmark set of 
parameters. We have indeed considered the pair production of a $\Sigma_\ell$ particle with 
an assignment of (10,400) GeV for $(m_\psi,m_{\Sigma_\ell})$ and fixed the couplings 
to give the correct relic density and a lifetime compatible with experimental limits. 
As evident from the results reported in table~(\ref{tab:ew_benchmark1}) the number of expected 
events is drastically reduced with respect to the colored case. A statistically relevant 
number of decay events both in the inner and outer part of the detector appears only at 
luminosities above $300\;{\mbox{fb}}^{-1}$. An eventual discovery of this scenario will thus 
require more time with respect to the case of color charged scalar field. We notice moreover 
that the low production cross section, combined with the low branching ratio $\sim 10^{-3}$, 
does not allow for any event with $\Sigma_\ell$ decaying into DM, even considering a high 
luminosity upgrade of the LHC. Since our choice of the mass of $\Sigma_\ell $ corresponds to 
the maximal cross-section available, it appears evident that, in the case of  $\Sigma_f$ charged 
only with respect to electroweak interactions, it will not be possible to identify at LHC the 
peculiar feature of our model, namely the presence of two (i.e. DM+SM and SM only) decay channels.

\begin{table}
\begin{center}
\begin{tabular}{|c|c|c|c|}
\hline
Part of detector & Total & $\Sigma \rightarrow DM$ & $\Sigma \rightarrow \mbox{SM only}$ \\
\hline
$\mathcal{L}=25 {\mbox{fb}}^{-1}$\\
\hline
Pixel & 1 & 0 & 1 \\
\hline
Tracker & 2 & 0 & 2 \\
\hline
Out & 23 & 0 & 23 \\
\hline
$\mathcal{L}=300 {\mbox{fb}}^{-1}$\\
\hline
Pixel & 14 & 0 & 14 \\
\hline
Tracker & 26 & 0 & 26 \\
\hline
Out &  270 & 0 & 270 \\
\hline
$\mathcal{L}=3000 {\mbox{fb}}^{-1}$\\
\hline
Pixel & 135 & 0 & 135 \\
\hline
Tracker & 260 & 0 & 260 \\
\hline
Out & 2705 & 0 & 2705 \\
\hline
\end{tabular}
\caption{Number of decay events, total as well as separately in the two kind of decay 
channels (DM or SM only), expected in the three detection regions at the LHC for the indicated 
values of luminosity. The benchmark chosen consist in a $\Sigma_\ell$-type field with mass of 
400 GeV while the  DM mass is set to 10 GeV. The pair ($\lambda,\lambda^{'}$) has been fixed 
to ($5\times 10^{-12},10^{-9}$).} 
\label{tab:ew_benchmark1}
\end{center}
\end{table}


Analogously to the colored scenario we can consider signals outside the double detection 
regions. In the high lifetime region only the decay channel into SM particles might be 
accessible to possible searches of stopped particles since the small number of events, 
consequence of the low production cross section, does not allow to have a statistically relevant 
number of events in both the decay channels~\footnote{We also notice that there are not, at the 
moment, searches of only electroweakly interacting stopped particles. On general grounds we 
expect a lower detection efficiency with respect to the case of color interaction.}. 
Alternatively we can consider very low values of the DM mass in order to evade ID constraints 
and for $x \sim 10^{-6}$ it might be again possible to observe both the $\Sigma_{\ell}$ decay 
channels, renouncing however to the possibility of next future detection of DM decays in cosmic 
rays. A possible outcome in this kind of scenario is reported in tab.~(\ref{tab:ew_benchmark2}), 
where it is shown the number of expected observed decay events in the case of $\Sigma_\ell$-type 
field with mass of 800 GeV and for a very low value of the DM mass corresponding to $x=10^{-6}$. 
The value of the $(\lambda,\lambda^{'})$ pair, respectively $8 \times 10^{-10}$, 
$7.8\times 10^{-10}$, guarantees the correct DM relic density, through the freeze-in mechanism, 
and a branching fraction of decay of the scalar field into DM of approximately $50\%$. Although 
the choice of the parameters is substantially analogous to the low mass benchmark studied in 
the case of color charged scalar (the lifetimes of the scalar field differ approximately by a 
factor 2) the majority of the events now lie in the ``outside'' region. This is due to the fact 
the $\Sigma_{\ell,e}$ are more boosted with respect to colored scalars. At the higher luminosities 
we have anyway a number of events in the inner detector, for both decay channels, satisfying our 
discovery requirement. We have thus again a good capability for LHC of providing information 
on the model under consideration, renouncing however at any prospects of indirect detection 
of the decays of the DM.   

\begin{table}
\begin{center}
\begin{tabular}{|c|c|c|c|}
\hline
Part of detector & Total & $\Sigma \rightarrow DM$ & $\Sigma \rightarrow \mbox{SM only}$ \\
\hline
$\mathcal{L}=25 {\mbox{fb}}^{-1}$\\
\hline
Pixel & 1 & 1 & 0 \\
\hline
Tracker & 2 & 1 & 1 \\
\hline
Out & 8 & 4 & 4 \\
\hline
$\mathcal{L}=300 {\mbox{fb}}^{-1}$\\
\hline
Pixel & 12 & 6 & 6 \\
\hline
Tracker & 27 & 14 & 13 \\
\hline
Out &  94 & 48 & 46 \\
\hline
$\mathcal{L}=3000 {\mbox{fb}}^{-1}$\\
\hline
Pixel & 120 & 62 & 58 \\
\hline
Tracker & 270 & 139 & 131 \\
\hline
Out & 941 & 484 & 457 \\
\hline
\end{tabular}
\caption{Number of decay events, total as well as separately in the two kind of decay 
channels (DM or SM only), which is expected in the three detection regions  at the LHC 
for the indicated values of luminosity. The benchmark chosen consist in a 
$\Sigma_\ell $-type field with mass of 800 GeV while the DM mass is set to 
$8 \times 10^{-4}$ GeV. The pair ($\lambda,\lambda^{'}$) has been fixed to 
($8\times 10^{-10},7.8\times 10^{-10}$) and corresponds to a lifetime many orders 
of magnitude above current and next future experimental sensitivity. No ID detection 
is thus expected in this case.} 
\label{tab:ew_benchmark2}
\end{center}
\end{table}


We conclude this section describing how the LHC signals may result altered, in the 
$\Sigma_\ell$ scenario, in the presence of a sizable mass splitting in the EW multiplet. 
As already mentioned in this case the heaviest component of the doublet dominantly decays 
into the lighter one and a W boson (or into two quarks/leptons) and these decays result 
prompt for mass-splitting above $\sim 1$ GeV. Contrary to the colored case, we can have 
a sizable production of the heavy component through the process 
$pp \rightarrow \Sigma_l^{\pm}\Sigma_l^{0}$, being mediated by the $W$ boson, which can 
account for $\sim 50\%$ of the total production cross-section even for sizable mass splittings. 
The range of signals results therefore enriched if the detection of the products of the 
prompt decays of the heavy component of the doublet is possible.   

In the case $m_{\Sigma^{\pm}} > m_{\Sigma_0}$ the decays $\Sigma^{\pm} \rightarrow \Sigma^{0} f 
\overline{f}^{'}$ produce displaced vertices accompanied by prompt leptons and jets, for 
short lifetimes of $\Sigma^0$, or a signal consisting of jet/leptons+missing energy, 
customarily studied in supersymmetric setups~\cite{Aad:2014vma,Khachatryan:2014qwa}, in the 
case that $\Sigma^{0}$ decays outside the inner detector~\footnote{Notice that we could as 
well consider monojet + missing energy signals in the case of long lived pair produced 
$\Sigma_l^0$, totally analogous to the DM pair production scenarios. However in our setup 
pair production of $\Sigma_l^{0}$ occurs only through a strongly off-shell Z-boson giving 
a rather weak signal~\cite{Fox:2011pm,CMS-PAS-EXO-12-048}.}.
In the opposite case $m_{\Sigma^{\pm}} < m_{\Sigma}^{0}$ prompt jets/leptons are associated 
to displaced vertices or metastable tracks.

\section{Discussion}

In this section we summarize the outcome of the previous analysis and discuss in greater 
detail whether an hypothetical next future LHC signal can unambiguously discriminate the 
underlying particle physics scenario. 

\begin{figure}[hbt]
\begin{center}
\subfloat{\includegraphics[width=6.8cm]{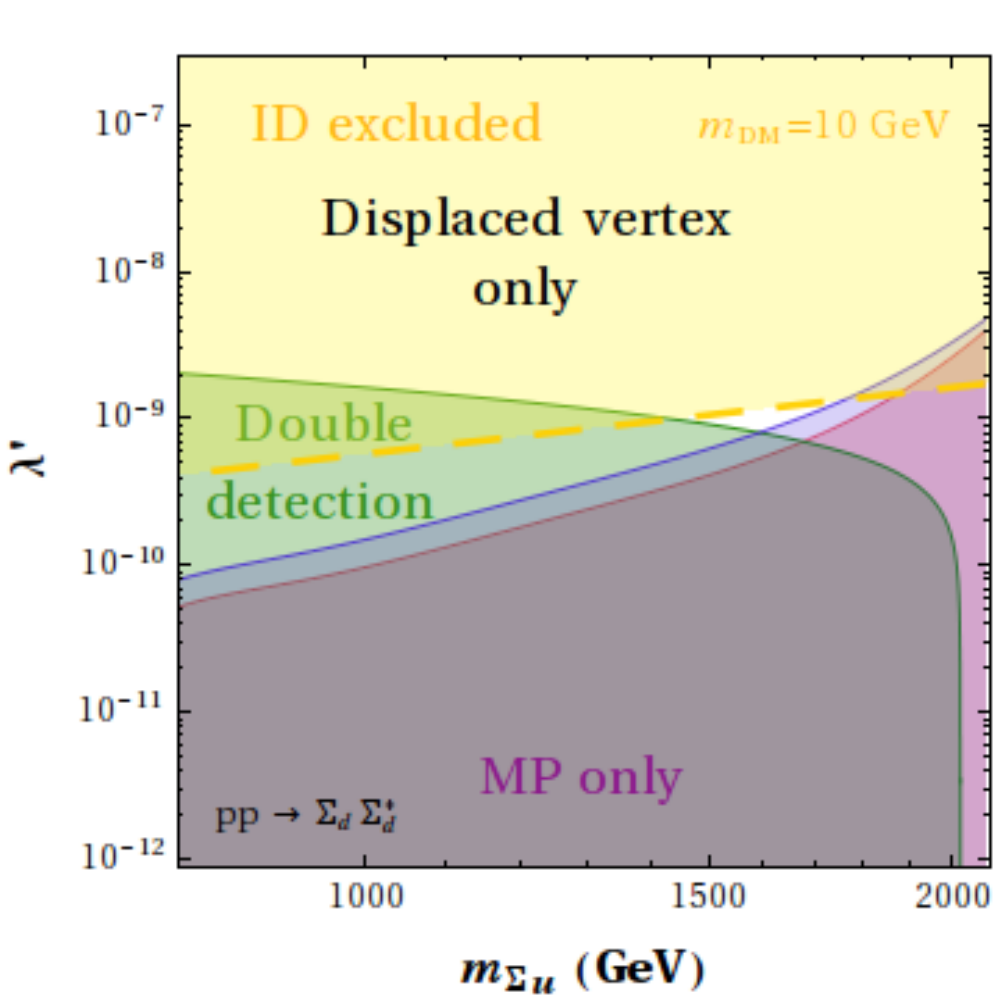}}
\hspace{10mm}
\subfloat{\includegraphics[width=6.8cm]{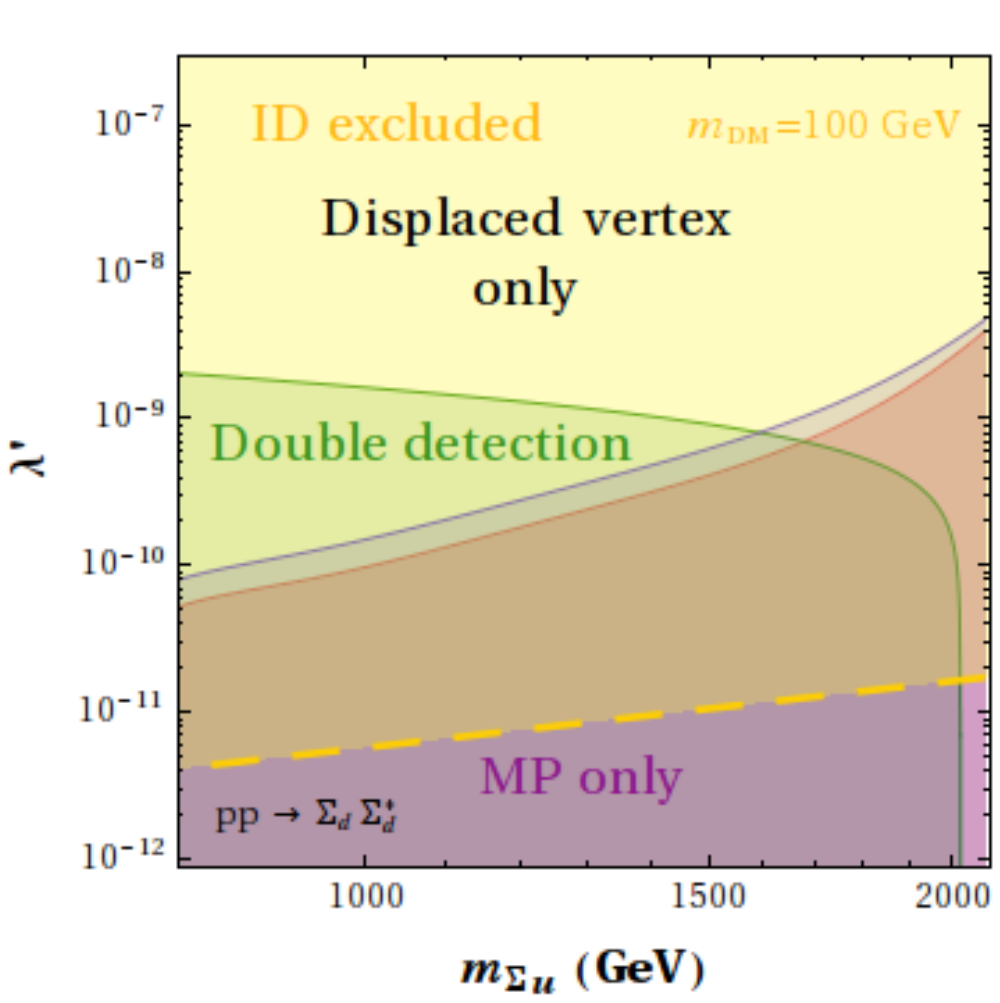}}
\caption{Summary of the possible observed signals at LHC, at $\mathcal{L}=300{\mbox{fb}}^{-1}$, 
for a $\Sigma_d$-type field, as function of its 
mass and the coupling $\lambda^{'}$. The other coupling, $\lambda$, has been fixed in order to 
reproduce the correct DM relic density while the DM mass has been set to 10 GeV (left plot) and 
100 GeV (right plot). The red, blue and green lines correspond to the observation of 10 events 
respectively in the Tracker, Pixel and outside the detector region. In the ``double'' 
detection region, the green shaded region between this lines, a number $\geq 10$ of decay 
events in the Pixel and the Tracker and more than 10 tracks leave the detector. The yellow 
shaded region, above the thick yellow dashed line, is excluded by constraints from indirect 
detection of DM decay.}
\label{fig:summary_colored}
\end{center}
\end{figure}

\begin{figure}[hbt]
\begin{center}
\subfloat{\includegraphics[width=6.8cm]{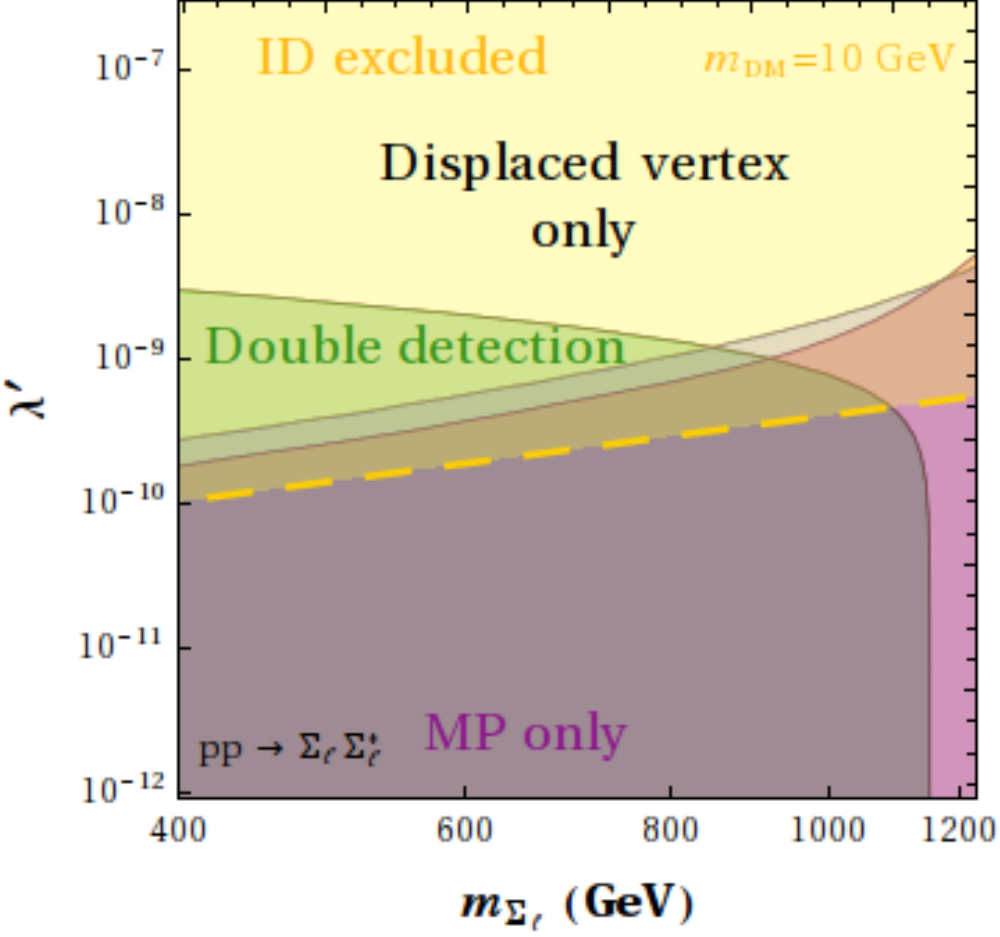}}
\hspace{10mm}
\subfloat{\includegraphics[width=6.8cm]{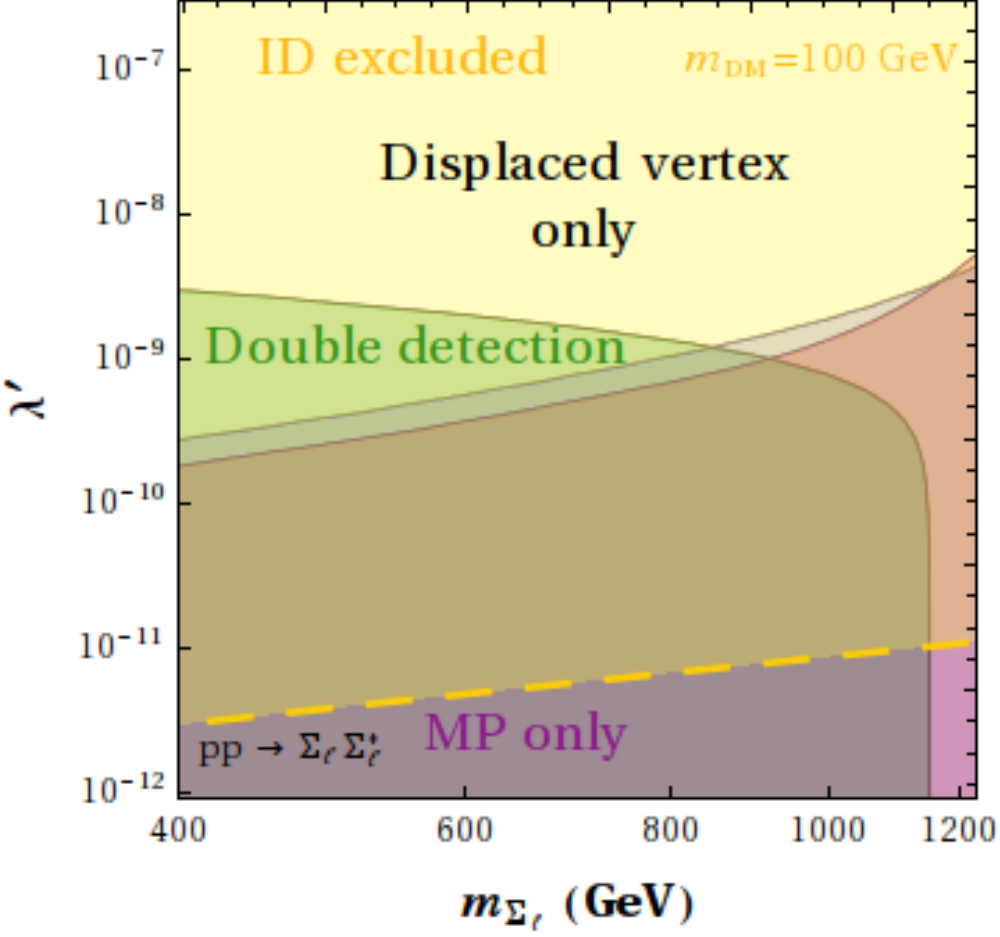}}
\caption{The same as fig.(\ref{fig:summary_colored}) but for the case of a $\Sigma_\ell$-type field.}
\label{fig:summary_ew}
\end{center}
\end{figure}

We show in fig.~(\ref{fig:summary_colored}) and~(\ref{fig:summary_ew}) the LHC reach, in 
the two scenarios of $\Sigma_d$- and $\Sigma_\ell $-type field, as function of the mass of 
the scalar field and of the coupling $\lambda^{'}$, for the two assignments 
of the DM mass of 10 and 100 GeV. The luminosity has been set to $300{\mbox{fb}}^{-1}$. 
The coupling $\lambda$ has been determined as function of the other parameters, using 
eq.~(\ref{Omegah2-DM2}), (\ref{eq:rate1}) and (\ref{eq:rate2}), according the requirement 
of the correct DM relic density.

The red, blue and green lines in the plots correspond to the observation of 10 events 
respectively in the Tracker, Pixel and outside the detector region. The green region comprised 
between this lines is the ``double detection'' region, in which the observation of at least 
10 events in the inner detector and at least 10 escaping tracks. Above this region only 
displaced vertices can be observed at a statistically relevant amount while below metastable 
tracks are the only visible signal. The yellow regions in the plots are already excluded, for 
the considered values of the DM mass, by constraints from DM indirect 
detection~\footnote{Notice that our bounds from ID detection are rather conservative. In the 
case of $\Sigma_{\ell,e}$-type field we can obtain, for example, weaker bounds assuming decays 
only in some flavour states like, e.g., $\tau$ leptons.}. As evident the possibility of 
``double'' LHC detection is already ruled out for DM masses above 100 GeV in the case of 
colored scalar and for masses above 10 GeV in the case of EW charged scalar. Nevertheless for 
the colored case, as can be seen in the left panel of fig.~(\ref{fig:summary_colored})
as well as in the benchmark in Table~1, the future indirect detection region just below the 
present bound lies exactly in the double detection corner for a DM mass of 10 GeV and 
$ m_{\Sigma} < 1500 $ GeV and in that case possibly all the four parameters of the model could 
be within reach in the next future. The possibility of observing only displaced vertices is 
disfavored as well by ID for these values of the DM mass, ad exception of the highest value of the 
scalar field mass, at the boundaries of the LHC reach. The more severe exclusion in this last 
case is due to the fact that the production of lighter scalar particles is considered, 
in turn implying higher decay rates for the DM.     

We can now investigate in more detail which kind of information can be inferred from an 
hypothetical future LHC signal. At high DM masses, namely above 100 (10) GeV in the colored 
(only electroweakly interacting) $\Sigma_f$ scenario, the only signal which can be expected 
are metastable tracks. In such case it will be possible to give a lower bound on the scalar 
field lifetime  or perhaps to measure it, although maybe with less precision with respect 
to a double detection scenario, from the decay of stopped $\Sigma_f$. From this it may be 
possible to determine the sum $\lambda^2+\lambda^{'\,2}$ while, in absence of observation 
of the decay processes, it is not possible to determine the single values of the two couplings, 
which could allow to infer the DM production mechanism. This task could be achieved in case 
of an Indirect Detection of dark matter decay which would provide, besides the value of the DM 
mass, information on a different combination of the two couplings, namely their product 
$\lambda \lambda^{'}$. The complete identification of the model could then be achieved by 
verifying that the DM relic density, computed with the reconstructed parameters, matches 
the cosmological value.

At intermediate DM mass scales the LHC ``double'' detection region is instead still viable. 
This scenario guarantees the optimal reconstruction of the lifetime of the scalar field, as 
well as its mass which can be inferred by the cross section and by the energy of the decay 
products. As already pointed out these informations alone do not allow to infer the DM relevant 
properties like the mass, which it is in any case not accessible at the LHC, and the production 
mechanism. Indeed the latter would require the knowledge of the single values of the couplings 
$\lambda$ and $\lambda^{'}$ while the lifetime of $\Sigma_f$ depends on the sum 
$\lambda^2+\lambda^{'\,2}$. The two couplings could be singularly inferred in case it is possible 
to distinguish, through the decay products in the displaced vertices, the two decay channels of 
the scalar field. However our study based on the relative branching fractions of the two decay 
channels of the scalar fields, confirmed from a quantitative perspective by the study of the 
two benchmarks reported in tab.~(\ref{tab:colored_benchmark}) and~(\ref{tab:ew_benchmark1}), 
has shown that only the pure SM decay channel is accessible to LHC detection. In this case it 
is possible to infer, from the determination of the lifetime of the scalar field, only the 
value of $\lambda^{'}$ (assuming a negligible branching fraction of decay into DM). 
A LHC signal only can thus neither provide evidence of the existence of the DM nor information 
on its production mechanism. A correlation with a DM ID signal is again mandatory for 
determining the remaining parameters.        
In both the scenarios proposed above the capability of full determination of the model under 
consideration is hence limited to the regions of the parameters space 
which lie in proximity of the current experimental sensitivity to DM Indirect Detection.

For very low DM masses, LHC is instead the only probe of the model under consideration while 
ID is not achievable (as a consequence, bounds like the ones shown in 
fig.~(\ref{fig:summary_colored}) and~(\ref{fig:summary_ew}), are completely evaded). On the 
contrary the two branching fractions of decay into the scalar field, namely the one into 
DM+SM and only SM, can be comparable within the double detection region. This statement is 
again confirmed at the quantitative level by the study of two benchmarks, one for colored and 
one for only EW interacting scalar, reported in tab.~(\ref{tab:colored_benchmark_2}) 
and~(\ref{tab:ew_benchmark2}), which show the presence of statistically relevant number of 
events, for both the two decay channels, in the inner detector as well as in the ``outside'' 
region. In case of identification of the two decay channels it is possible to infer the 
values of $\lambda$ and $\lambda^{'}$ as well as the $m_{\Sigma_f}$. The DM mass is instead 
not directly accessible from observations but might be determined from the requirement of 
the correct relic density, given its proportionality to $x=m_\psi/m_{\Sigma_f}$.  

We remark anyway that the capability of disentangling the decay channels of the scalar 
field is actually model dependent since, according to which of the operators 
in~(\ref{eq:DMint}) and (\ref{eq:lagrangian}) determine the decay processes of $\Sigma_f$,
we might have scenarios with very different decay products in the two channels (for example 
in the case of $\Sigma_\ell$ field we can have a decay into DM and a charged lepton opposed to 
a decay into two jets triggered by the coupling $\lambda_{2,3\,\ell}$) as well as rather similar 
signals, e.g. in the case in which the scalar field decays into a neutrino and another SM 
fermion. We also remark that the various kinds of final states, e.g. jets, leptons etc..., 
possibly emerging from displaced vertices have not the same capability of reconstruction, 
at a given lifetime~(see e.g.~\cite{Graham:2012th} for a discussion). In order to properly 
address this issue a refinement of our analysis, including a simulation of the 
detector, is required.

The statements discussed until now substantially hold for all the assignments of the 
SM charge of the $\Sigma_f$ particle. In absence of a detector analysis, sensitive to the 
decay products, the main difference is between the cases of colored or only electroweakly 
charged $\Sigma_f$ with the latter requiring higher luminosities for a discovery and featuring 
a more limited mass reach, in view of the lower production cross-section.  
In the case of colored mediator one could consider complementary searches, with respect to 
the one discussed in this paper, relying on the possibility of observing very late decays of 
$\Sigma_f$ particles stopped in the detector. At these very long lifetimes the observation of 
both decay channels of $\Sigma_f$ is compatible with constraints from DM phenomenology. Our 
study reported in tab.~(\ref{tab:colored_benchmark_1}) shows that this possibility is potentially 
feasible even once the actual low efficiencies in this kind of searches are accounted for. 
On the other hand we remark again that in order to properly determine whether the different 
decay channels can be discriminated a simulation of the detector is mandatory and that a 
definite statement is left to a future study. In case of possible identifications of the 
two decay channels it is again possible to determine their relative branching fractions and 
then directly the couplings $\lambda$ and $\lambda^{'}$.

\section{Conclusions}

We have considered the LHC detection prospects of a very simple and economical extension 
of the Standard Model featuring a decaying Maiorana fermion as DM candidate and a 
metastable scalar field with non-trivial quantum numbers under the SM gauge group. The potential 
LHC signal is constituted by the observation of the displaced decays of the scalar field, either 
into the DM and a SM fermion or into two SM fermions, or of the scalar field itself as 
disappearing track escaping the detector. We have investigated the possibility of detecting 
this kind of signals in the regions of the parameters space favored by the requirement of 
the correct DM relic density and a future Indirect Detection of the DM decays.

In the case of color charged scalar field a statistically relevant population of both 
displaced vertices and metastable tracks might be observed in next future 14 TeV LHC run up 
to masses of the order of 1500 GeV and up to approximately 2200 considering a high luminosity 
run; however bounds from DM Indirect Detection already exclude the possibility of "double" 
detection for DM masses above 100 GeV, leaving open only a possible detection of metastable tracks.
In the parameter region corresponding to the ``double'' LHC detection unfortunately only 
the decay channel into just SM fermions is accessible to observation, since the branching 
ratio of the scalar field into DM is of the order $ \leq10^{-2}-10^{-3}$. The two decay 
channels of the $\Sigma_f$ field might be instead contemporarily observed in the case of 
very long $\Sigma_f$ lifetimes, with negligible amount of events in the inner detector, 
considering the possibility that a fraction of $\Sigma_f$ can be stopped in the detector 
material and decay at later time. This possibility should be investigated through a dedicated 
study. The contemporary detection of the two decay channels of $\Sigma_f$ can occur as 
well at very low, i.e. $ \ll 1\,\mbox{GeV}$ DM masses. In this case there would be the 
observation of mostly displaced vertices in the inner detector, while DM decays in Indirect 
Detection are unfortunately beyond present and future observational capabilities 
because of the extremely long DM lifetimes.

In the case in which, instead, $\Sigma_f$ features only electroweak interactions the maximal 
mass reach is limited to approximately 1400 TeV and the high luminosity upgrade is most 
likely needed to probe the parameters space of model. Because of the lower masses of 
$\Sigma_f$ involved in the analysis the bounds from DM ID are more severe, with respect 
to the case of a colored scalar field, such that under the most conservative assumptions the 
``double'' LHC detection is not possible for DM masses above 10 GeV. There are as well poorer 
prospects of detection of stopped particles because of the lower number of events and of 
stopping probability. The scenario of contemporary detection of the two decay channels at 
very low DM masses is instead feasible although kinematic effects imply, contrary to the 
colored case, a greater number of decays in the outer detector region. The range of possible 
signals is enriched in the case the scalar field is a $SU(2)_L$ doublet with a mass splitting 
between the components. In particular, in the case of $\Sigma_\ell $ type field, with lighter 
neutral component, the signals presented above can be accompanied as well by conventional 
missing energy signatures and prompt tracks.

In most of the parameter space of the scenario presented, LHC experiments and Indirect 
Detection observations provide complementary information on the model and both signals are 
necessary in order to confirm the identity of the Dark Matter and verify the production mechanism. 
In particular if the mass of the Dark Matter is in the 1-10 GeV range and the $\Sigma_f$ scalar 
within the kinematical reach of the next LHC run, i.e. below 1600 GeV for the colored case or 
1100 GeV for the EW case, there is a clear chance to obtain multiple signals in the very near future. 

\acknowledgments

The authors thank Marco Nardecchia for enlighting discussions.

\noindent
The authors acknowledge partial support from the European Union 
FP7 ITN-INVISIBLES (Marie Curie Actions, PITN-GA-2011-289442).

\bibliography{bibfile}{}
\bibliographystyle{JHEP}
\end{document}